\documentclass[paper,notoc,nohyper]{JHEP3}
\usepackage{epsfig}
\usepackage{amsbsy} 
\usepackage{graphics} 
\def\PL #1 #2 #3 {{\it Phys. Lett.} {\bf#1} (#3) #2}
\def\NP #1 #2 #3 {{\it Nucl. Phys.} {\bf#1} (#3) #2}
\def\ZP #1 #2 #3 {{\it Z. Phys.} {\bf#1} (#3) #2}
\def\PRL #1 #2 #3 {{\it Phys. Rev. Lett.} {\bf #1} (#3) #2}
\def\PR #1 #2 #3 {{\it Phys. Rev.} {\bf#1} (#3) #2}
\def\MPL #1 #2 #3 {{\it Mod. Phys. Lett.} {\bf#1} (#3) #2}
\def\RMP #1 #2 #3 {{\it Rev.~Mod. Phys.} {\bf#1} (#3) #2}

\newcommand{\Pp}[1]{\mbox{$\left( #1 \right)^2$}}
\newcommand{\e}{\epsilon}

\newcommand{\CC}{\,{\cal C}}
\newcommand{\DD}{\,{\cal D}}
\newcommand{\JJ}{{\cal S}}

\newcommand{\PP}{{\cal P}}
\newcommand{\TT}{{\cal T}}

\newcommand{\setqi}{\{q_i\}}
\newcommand{\setnui}{\{\nu_i\}}
\newcommand{\setone}{\{1\}}
\newcommand{\G}[1]{\Gamma\left({#1}\right)}

\newcommand{\halfD}{\frac{D}{2}}

\title{\boldmath 
 A Calculational Formalism for One-Loop Integrals
}

\author{
W.T.~Giele$^a$ and E.W.N.~Glover$^b$\\
$^a$Fermilab, 
Batavia, 
IL 60510, USA\\
$^b$Department of Physics, 
University of Durham, 
Durham DH1 3LE, 
England\\ 
E-mail:  \email{giele@fnal.gov}, \email{E.W.N.Glover@durham.ac.uk}}
\abstract{
We construct a specific formalism for calculating the one-loop
virtual corrections for standard model processes with
an arbitrary number of external legs.
The procedure explicitly separates the infrared and ultraviolet 
divergences analytically
from the finite one-loop contributions, which can then be evaluated 
numerically using recursion
relations.
Using the formalism outlined in this paper, we are in position
to construct the next-to-leading order corrections to a variety of
multi-leg QCD processes such as multi-jet production and vector-boson(s)
plus multi-jet production at hadron colliders.
The final limiting factor on the number of particles will
be the available computer power.  
}

\preprint{{FERMILAB-PUB-04/021-T}, {DCTP/04/08}, {IPPP/04/04}, {hep-ph/0402152}}

\begin{document}

\section{Introduction}
Multi-jet and vector-boson(s) plus multi-jet events are already common 
at the
TEVATRON and will become even more frequent at the LHC.   In addition to the
information they contain about the strong dynamics, they also form important
backgrounds to new physics. At present, the rates for such processes can be
evaluated relatively easily at leading order (LO) in parton level 
perturbation
theory~\cite{LOrecur,LOdiagram}.   However, the predicted event rate depends
strongly on the choice of the coupling constant (or equivalently the
renormalisation scale) so that the calculated rate is only an ``order of 
magnitude''
estimate. In addition, there is a rather poor mismatch between the ``single
parton becomes a jet"  approach used in LO perturbation theory and the
complicated multi-hadron jet observed in experiment.   While these problems
cannot be entirely solved within perturbation theory, the situation can be
ameliorated by calculating the strong next-to-leading order (NLO) 
corrections.
This reduces the scale dependence  and the additional 
parton radiated into the final state allows a better
modelling of the inter- and intra-jet energy flow as well as identifying regions
where large logarithms must be resummed.

While many NLO calculations are available for $2\to 2$ processes in hadron
colliders\footnote{Here, attempts are currently underway to evaluate the
next-to-next-to-leading order (NNLO) corrections with the aim of 
reducing the
theoretical uncertainty to the level of a few  per cent - the anticipated
experimental accuracy. See Ref.~\cite{NNLO} for an overview.}, 
relatively few
estimates of $2\to 3$ processes based on five-point one-loop and 
six-point tree
matrix elements exist. Currently, numerical programs capable of 
producing NLO
predictions for fully differential distributions are available for
3-jet~\cite{3jet} and V+2-jet~\cite{V2jet} final states, as well as the
important  $pp \to t\bar tH$~\cite{ttH} and $pp \to H$+2-jet~\cite{H2jet}
signatures of the Higgs boson.  The NLO predictions for important background
processes such as $pp \to VV$+jet and $pp \to Q\bar Q$+jet are not available
yet.\footnote{Some of the same five parton one-loop matrix elements~\cite{V5par} have
been used to predict four jet event shapes in electron-positron
annihilation~\cite{ee4jet} as well as $3+1$-jet rate in electron-proton
collisions~\cite{dis3jet}.}

The ingredients necessary for computing
the NLO correction to a $N$ particle process are well
known.   First one needs the tree-level contribution for the $N+1$ particle
process where an additional parton is radiated.  Second, one needs the
one-loop  $N$ particle matrix elements.  Both terms are infrared (and 
usually
ultraviolet) divergent and must be carefully combined to yield an 
infrared
and ultraviolet finite NLO prediction.    

The radiative contribution is relatively well under control and can 
easily be
automated.  Computer programs exist that can generate the matrix
elements (and associated phase space) either Feynman diagram by Feynman
diagram~\cite{LOdiagram} or using recursion relations~\cite{LOrecur}.  
The infrared singularities that occur when a parton is soft (or when
two partons become
collinear) can then be removed using well established (dimensional 
regularisation)
techniques so that the ``subtracted" matrix element is finite
and can be evaluated in 4-dimensions~\cite{treesub}.    

In contrast, the one-loop contribution is usually obtained analytically on a
case by case basis and extensive computer algebra is employed to extract the
infrared and ultraviolet singularities in one-loop graphs that appear as 
poles
in the dimensional regularisation parameter $\epsilon = (4-D)/2$ and 
then to evaluate
the finite remainder in terms of logarithms and dilogarithms. 
At present, this is the
bottleneck for producing NLO corrections for a variety of multiparticle
processes.

The NLO corrections to the $2\to 3$ processes described above require 
one-loop
calculations involving up to 5 external particles, i.e. the one-loop 
five-point function.
The limiting factor is the number of external  particles 
involved.
While the one-loop $N$-point function can recursively be expressed in
$(N-1)$-point functions all the way down to simple three-point
functions \cite{binoth}, the algebraic complexity as $N$ increases quickly
becomes overwhelming.\footnote{The fact 
that the one-loop $N$-point function can recursively be expressed 
in $(N-1)$-point functions was well known for 4-dimensional integrals~\cite{looprecur}.
However, the extension to dimensionally regulated integrals was not
obvious as the identities were based on the fact that space-time is
four dimensional. The extension to higher dimensions was first formalized
in Refs.~\cite{looprecur1,Abox} and further developed by
a variety of authors~\cite{looprecur2,binoth,Nizic}.
} This is mainly due to the rapidly increasing number of
kinematic scales present in the problem.

Adding one further particle to the final state corresponds to $2 \to 4$ 
hadron
collider processes.  The NLO corrections to such processes requires the
knowledge of one-loop six-point functions.   To date, no standard model
computations  of one-loop six-point amplitudes exist.\footnote{Note that 
the six-point
amplitudes have been evaluated in the Yukawa model~\cite{binoth6,leshouches}.}

Rather than evaluating the one-loop matrix elements algebraically, an
alternative  would be a numerical approach~\cite{formf,passarino,binothsector}. 
There are two main problems.  First, the ultraviolet and infrared singularities 
must be handled correctly  so 
that
they can be cancelled against those coming from the radiative contribution.
Second, when anomalous thresholds occur in the loop integral,
special attention must be given to the behaviour around (possibly) singular
points in phase space~\cite{passarino}.

Nevertheless, several avenues of attack are readily apparent. One 
approach that
has already been successfully used for calculating five-point matrix 
elements
would be to construct the loop amplitude by ``sewing" tree amplitudes
together~\cite{cut}. Another way would be to combine the virtual and real
contributions so that the divergences cancel directly in the integration 
over
the loop momentum~\cite{soper}. Alternatively,  one can construct 
directly the
counterterm diagram by diagram~\cite{sopernagy}.  

Here, we follow a different path. It is well known that (tensor) one-loop
integrals can be written as combinations of finite four-point scalar 
integrals
in $D+2$ dimensions  and  infrared/ultraviolet divergent triangle graphs in
$D$ dimensions~\cite{looprecur,binoth,Nizic}.   
For example,  in Ref.~\cite{kauer} the scalar six-point
function is analytically expressed in terms of $D$ dimensional triangles and
$D+2$ dimensional boxes using the recursive techniques of 
Ref.~\cite{binoth}.
Using this result one can explicitly separate the divergent 
contributions, which
are computed analytically in $D$ dimensions using dimensional 
regularization,
from the finite contributions which can be evaluated numerically in 4
dimensions.  However, using  the recursion relations to calculate the
divergent part  of multi-leg loop integrals leads to an overwhelming
algebraic complexity, particularly when $N \geq 6$ and when  more kinematic
scales  are involved in the problem. While the 
infrared pole structure is always very simple, each level of recursion
introduces additional inverse powers of determinants of kinematic matrices which
are difficult to cancel analytically.
In this paper we advocate a different
technique to directly extract the soft/collinear divergences from the loop
graph and express them in the basis set of divergent $D$ dimensional 
triangle
graphs\footnote{Note that a similar approach has been employed by
Dittmaier~\cite{dittmaier}. Although our method is very similar for the 
scalar
integrals, it differs in the treatment of the tensor integrals.}. 
This approach
to extract the divergent contribution allows us to minimize the algebraic
manipulations needed to isolate the singularities.
Once the coefficients of the divergent integrals
are known, we can eliminate the divergent triangles from the basis set of
integrals, leaving us only with a basis of (known) finite integrals. The
kinematic coefficients multiplying the finite integrals can then be 
numerically
evaluated using the recursion relations
thereby maximizing the number of external legs that can be handled. 

Our paper is constructed as follows.  First, in Sec.~\ref{sec:notation}  we
define the necessary formalism for  relating tensor integrals in 
$D\sim
4$ dimensions to integrals in higher dimensions with additional powers 
of the
propagators.  We discuss the infrared and ultraviolet divergence 
conditions for
these integrals and establish how to isolate the divergences using a set of
augmented recursion relations based on the results of 
ref.~\cite{binoth}. The
basis set that emerges is quite natural, divergent triangle functions and a
collection of (finite) higher dimension box functions. We show how the
recursion relations can be used to numerically evaluate the set of finite
integrals.   

In Sec.~\ref{sec:subtract}, we show how to determine the singular 
contributions
directly without use of the recursion relations. For simplicity we work 
in the
limit where all of the internal particles are lightlike and construct  a 
master
formula for the divergent term using  divergent triangles as the basis 
set of
functions.    The coefficients of each of the divergent
three-point function is determined in
Sec.~\ref{subsec:c123}.
Our conclusions are summarized in Sec.~\ref{sec:conclusions}.

Several technical appendices have been added. 
Appendix~\ref{app:recursion} gives details about the derivation of the
basic recursion relations together with explicit construction
prescriptions for the numerical calculation of the kinematic
coefficients appearing in the recursion relations.
In Appendix~\ref{app:analytic}, we list analytic expressions
for the basis sets of divergent integrals that appear
once the recursion relations have been applied
are specified.
Finally, Appendix~\ref{app:scalar} contains an explicit expression for the
soft/collinear contributions from an on-shell massless $N$-point function. 

\section{Outline of the Formalism}
\label{sec:notation}

In this section we give an explicit outline for calculating one-loop matrix
elements with an arbitrary number of external legs. The strategy is to reduce
the $N$-point one-loop integrals to a basis set of calculable master integrals.

We start in Sec.~\ref{sec:decomposition} by  relating the tensor
integrals appearing in dimensionally regulated Feynman diagram calculations
(in  $D = 4-2\epsilon$ dimensions) to integrals in higher dimensions with
additional powers  of the propagators. This enlarged set of integrals
naturally breaks down into three classes of scalar integrals: finite,
ultraviolet (UV) divergent and infrared (IR) divergent integrals.  We
discuss the infrared and ultraviolet divergence  conditions for one-loop
integrals in arbitrary dimension and with arbitrary powers of the
propagators in Sec.~\ref{sec:classification} and develop the necessary
basis sets for the divergent integrals. 
In Sec.~\ref{sec:recursion} we derive a complete  
system of augmented recursion relations needed for the formalism.
Finally, in Sec.~\ref{sec:basis}, we formulate the
decomposition of the amplitude into a set of master
integrals produced by the recursion relation.

\subsection{Decomposition of One-Loop Graphs to Master Integrals}
\label{sec:decomposition}

The one-loop $M$-particle amplitude ${\cal A}_M$ with external momenta $p_i$
can be written as a sum over the
individual one-loop Feynman graphs ${\cal M}_G$, 
\begin{equation} 
{\cal A}_M(p_1,p_2,\ldots,p_M)=\sum_G {\cal M}_G(p_1,p_2,\ldots,p_M).
\end{equation} 
Each of the Feynman graphs
is itself a sum over different rank-$m$ $N$-point functions, where $m\leq N\leq M$,
\begin{eqnarray} 
{\cal M}_G(p_1,p_2,\ldots,p_M)&=&\sum_G
\sum_{m=0}^{N} {\cal C}^{G}_{\mu_1\mu_2\cdots\mu_m}
I_N^{\mu_1\mu_2\cdots\mu_m}(D;\setqi,\setone).
\end{eqnarray}
The coefficient ${\cal C}^G_{\mu_1\mu_2\cdots\mu_m}$ is typically composed of
tree-level multi-particle currents that depend on the properties of the
external particles (such as spin, polarization and ultimately on the momenta).
In general this coefficient can depend on dimensional factors (i.e. it
depends on the regulator $\e$). 
Here, we are more concerned with the rank-$m$  $N$-point tensor integrals with unit propagators
that are defined as
\begin{equation}
\label{eq:masterintegral}
I_N^{\mu_1\mu_2\cdots\mu_m}(D;\setqi,\setone)= \int\frac{d^D \ell}{i\pi^{D/2}}
\frac{\ell^{\mu_1}\ell^{\mu_2}\cdots \ell^{\mu_m}}{d_1d_2\cdots d_N},
\end{equation}
where the propagator terms are given by $d_i=(\ell+q_i)^2+i0$. 

\FIGURE[t!]{
\label{fig:generic}
\scalebox{.5}{\includegraphics{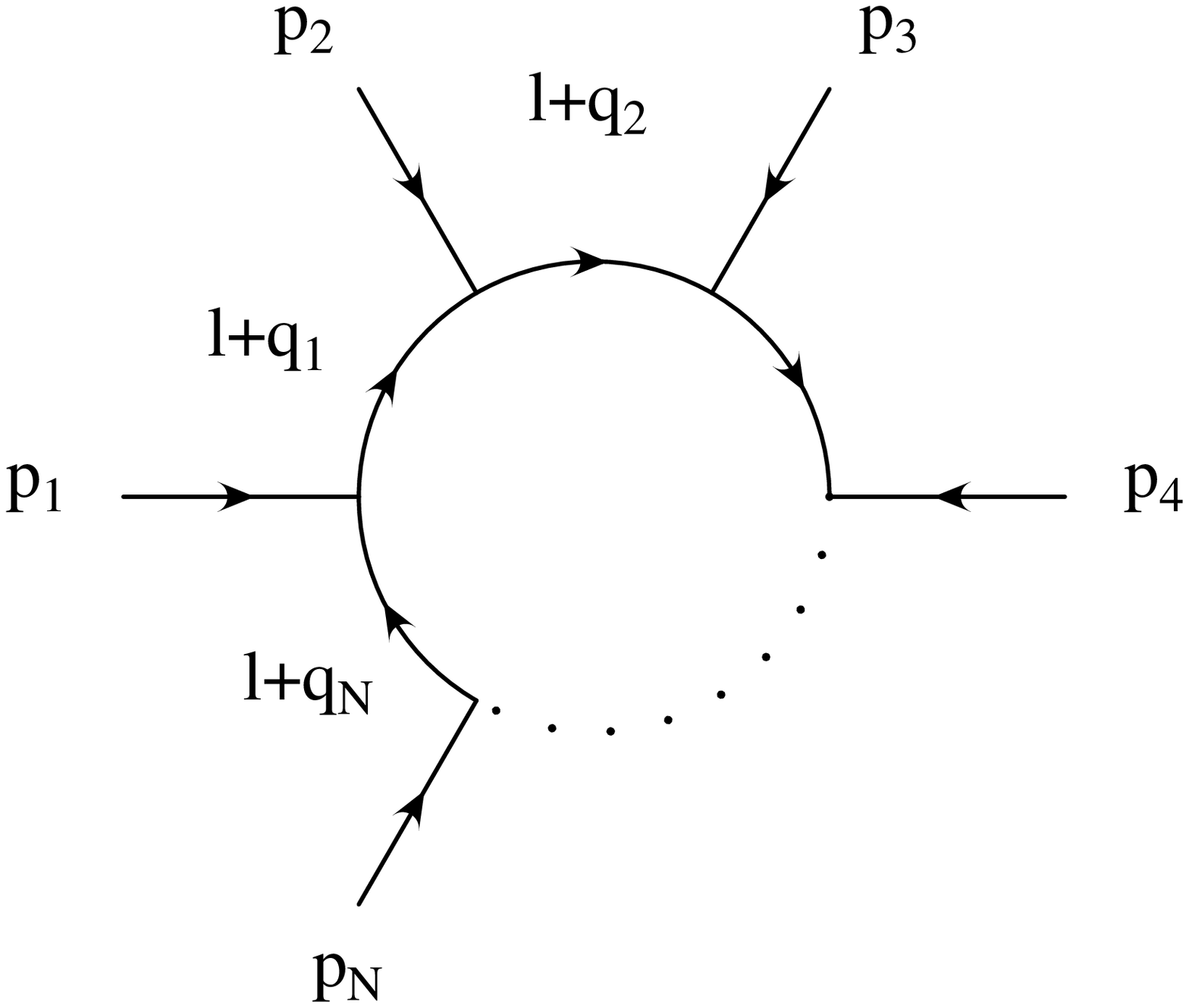}}
\caption{The generic $N$-point loop graph.}
}

As can be seen in Fig.~\ref{fig:generic}, the momenta $\setqi$ that characterise the
loop integral are composed from the external momenta. Depending on the
topology of the diagram the $q_i$ are sums over different $p_i$.  The notation $\setone$
indicates that all of the propagators are raised to unit powers.

The generic tensor integral of Eq.~(\ref{eq:masterintegral})  
contains three different classes
of divergences. The first class contains the so-called rescattering 
singularities. That is, two non-adjacent propagators are on-shell. These
singularities are protected by the $+i0$ prescription and generate the complex
part of the integral. In an analytic calculation these contributions are
generated by analytic continuations of the  transcendental functions into the
physical region. 
The second class of singularities are generated when two adjacent  propagators
become singular.
These are the genuine soft/collinear singularities.  
The final class of singularities occurs when the loop momentum
becomes large - the ultraviolet singularities.  

To reduce the tensor integral to scalar integrals we use the
method developed in Ref.~\cite{tensor} to extract the Lorentz structure  
and raise both the dimension and powers of the propagators within the remaining scalar
loop integral such that
\begin{eqnarray}
\label{eq:davydychev}
I_N^{\mu_1\mu_2\cdots\mu_m}(D;\setqi,\setone)&=&
\sum_{\lambda,x_1,x_2,\ldots,x_N}\delta_{(2\lambda+\sum_i x_i-m)}
\left(-\frac{1}{2}\right)^\lambda x_1!x_2!\cdots x_N!
\nonumber\\
&\times&\left\{g^\lambda q_1^{x_1}q_2^{x_2}\cdots
q_N^{x_N}\right\}^{\mu_1\mu_2\cdots\mu_m}
\nonumber\\&\times&
I_N(D+2(m-\lambda);\setqi,\{1+x_i\}).
\end{eqnarray}
The structure
$\left\{g^\lambda q_1^{x_1}q_2^{x_2}\cdots q_N^{x_N}\right\}^{\mu_1\mu_2\cdots\mu_m}$
means we assign the Lorentz indices in all distinct manners to a number of $\lambda$ metric tensors
$g_{\alpha\beta}$, $x_1$ momentum vectors $q_1$, etc. For example,
\begin{eqnarray}
\{q_1q_1\}^{\mu_1\mu_2}&=&q_1^{\mu_1}q_1^{\mu_2},
\nonumber\\
\{q_1q_2\}^{\mu_1\mu_2}&=&q_1^{\mu_1}q_2^{\mu_2}+q_1^{\mu_2}q_2^{\mu_1},
\nonumber\\
\{gq_1\}^{\mu_1\mu_2\mu_3}&=&g^{\mu_1\mu_2}q_1^{\mu_3}+g^{\mu_2\mu_3}q_1^{\mu_1}
                           +g^{\mu_3\mu_1}q_1^{\mu_2},
\nonumber\\
\{q_1^2q_2^2\}^{\mu_1\mu_2\mu_3\mu_4}&=&q_1^{\mu_1}q_1^{\mu_2}q_2^{\mu_3}q_2^{\mu_4}
+q_1^{\mu_1}q_1^{\mu_3}q_2^{\mu_2}q_2^{\mu_4}
+q_1^{\mu_1}q_1^{\mu_4}q_2^{\mu_2}q_2^{\mu_3}
\nonumber\\
&+&q_1^{\mu_2}q_1^{\mu_3}q_2^{\mu_1}q_2^{\mu_4}
+q_1^{\mu_2}q_1^{\mu_4}q_2^{\mu_1}q_2^{\mu_3}
+q_1^{\mu_3}q_1^{\mu_4}q_2^{\mu_1}q_2^{\mu_2}.
\end{eqnarray}
Finally, the generalised $N$-point scalar integral with raised propagator powers
is defined by
\begin{equation}
I_N(D;\setqi,\setnui)=
\int\frac{d^D \ell}{i\pi^{D/2}}\frac{1}{d_1^{\nu_1}d_2^{\nu_2}\cdots d_N^{\nu_N}}.
\end{equation}
As we will see later, it is often convenient to classify the integral 
according to the value of 
\begin{equation}
\label{eq:sigmadef}
\sigma=\nu_1+\nu_2+\cdots+\nu_N.
\end{equation}
Note that if one of the $\nu_i=0$ the corresponding propagator 
is removed yielding a ($N-1$)-point function.
Furthermore, for notational purposes we will often suppress
the momentum arguments\footnote{We remind the reader that when a propagator is 
pinched out, and the topology of the loop integral is changed, 
the momenta $\setqi$
associated with that particular graph also changes.}
\begin{equation}
I_N(D;\setqi,\setnui) \equiv I_N(D;\{\nu_l\}).
\end{equation}
Throughout the paper we will follow this notation which is similar to 
that developed in
Ref.~\cite{Nizic}.

Generically, these scalar integrals in higher dimensions and
with repeated propagators can be related using
recursion relations, to three basis sets of integrals that reflect the
singularity properties of the integrals: the set of finite integrals, ${\cal
I}^{fin}$,  the set of infrared divergent integrals, ${\cal I}^{IR}$ and the
set of ultraviolet divergent integrals,  ${\cal I}^{UV}$. The one-loop
amplitude can therefore be written as,
\begin{eqnarray}
\label{eq:ansatz}
{\cal A}_M(p_1,p_2,\ldots,p_M)&=&{\cal A}_M^{IR}(p_1,\ldots,p_M)
+{\cal A}_M^{fin}(p_1,\ldots,p_M)+{\cal A}_M^{UV}(p_1,\ldots,p_M)\nonumber\\
&=&\sum_i K_i^{IR}{\cal I}^{IR}_i(\{q_j\})+\sum_i K_i^{fin}{\cal I}^{fin}_i(\{q_j\})
+\sum_i K_i^{UV}{\cal I}^{UV}_i(\{q_j\}),\nonumber\\
\end{eqnarray}
where the summations run over the integrals in each of the three basis sets.
The kinematic factors $K$ are functions of the kinematic scales in the
process.
To make Eq.~(\ref{eq:ansatz}) more concrete, we first classify
each of the scalar integrals 
that can appear in Eq.~(\ref{eq:davydychev}) 
according to whether it is finite, IR divergent or UV divergent and then reduce 
it to the basis set using the appropriate recursion relations.

\subsection{Classification of Integrals}
\label{sec:classification}

The first important step is to identify the conditions under which a
specific integral is finite, IR divergent or UV divergent.\footnote{
In the following, we give a general discussion of the divergent conditions.
A more detailed analysis of the singularity structure of integrals with $N \leq 6$ 
 including a thorough discussion of the role of internal masses and
 anomalous thresholds is given in Ref.~\cite{passarino}.} 
Ultimately,
these conditions will control which recursion relation to use in the reduction towards
the master integrals.
The integral $I_N(D;\setnui)$ 
{\bf may} be infrared divergent only when
\begin{equation}
\label{eq:IRcond}
\sigma-N+2 = \frac{D}{2},
\end{equation}
depending on whether or not the external legs are on-shell
and {\bf is} ultraviolet divergent when
\begin{equation}
\label{eq:UVcond}
\sigma = \frac{D}{2}.
\end{equation}
$\sigma$ is defined in Eq.~(\ref{eq:sigmadef}) and, since $\sigma \geq N$, 
integrals may be finite
even if the external legs are on-shell provided that $D > 4$ and the condition
\begin{equation}
\label{eq:fin}
  \sigma-N+2 < \frac{D}{2} < \sigma  ,
\end{equation}
is satisfied.  When an integral is IR and/or UV divergent, 
we wish to extract the divergences 
 and relate the finite remainder to the set of 
 finite integrals using recursion relations.
\FIGURE[t!]{
\label{fig:Dsigma}
   (a)
\scalebox{.3}{\includegraphics{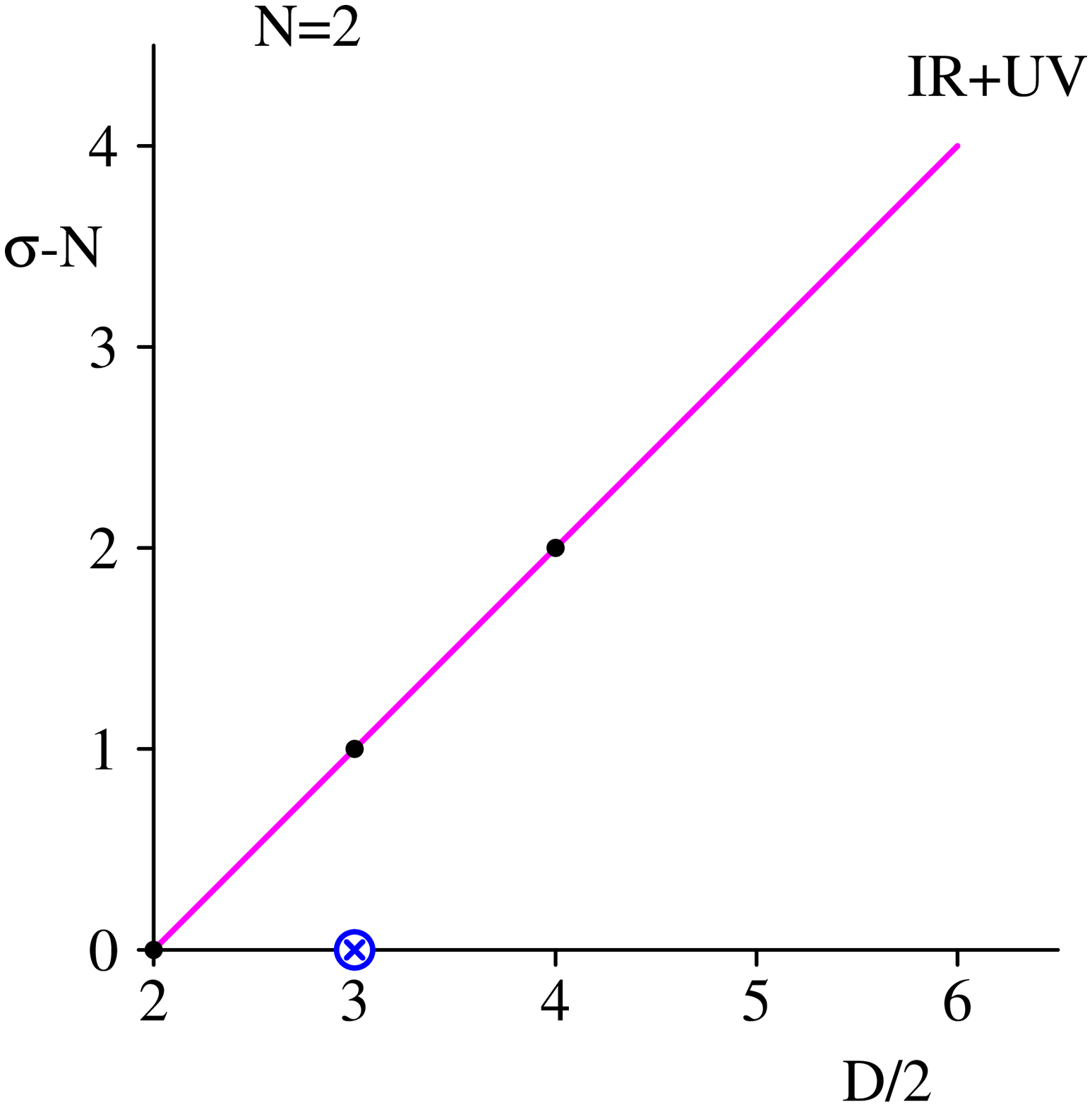}}
   (b)
\scalebox{.3}{\includegraphics{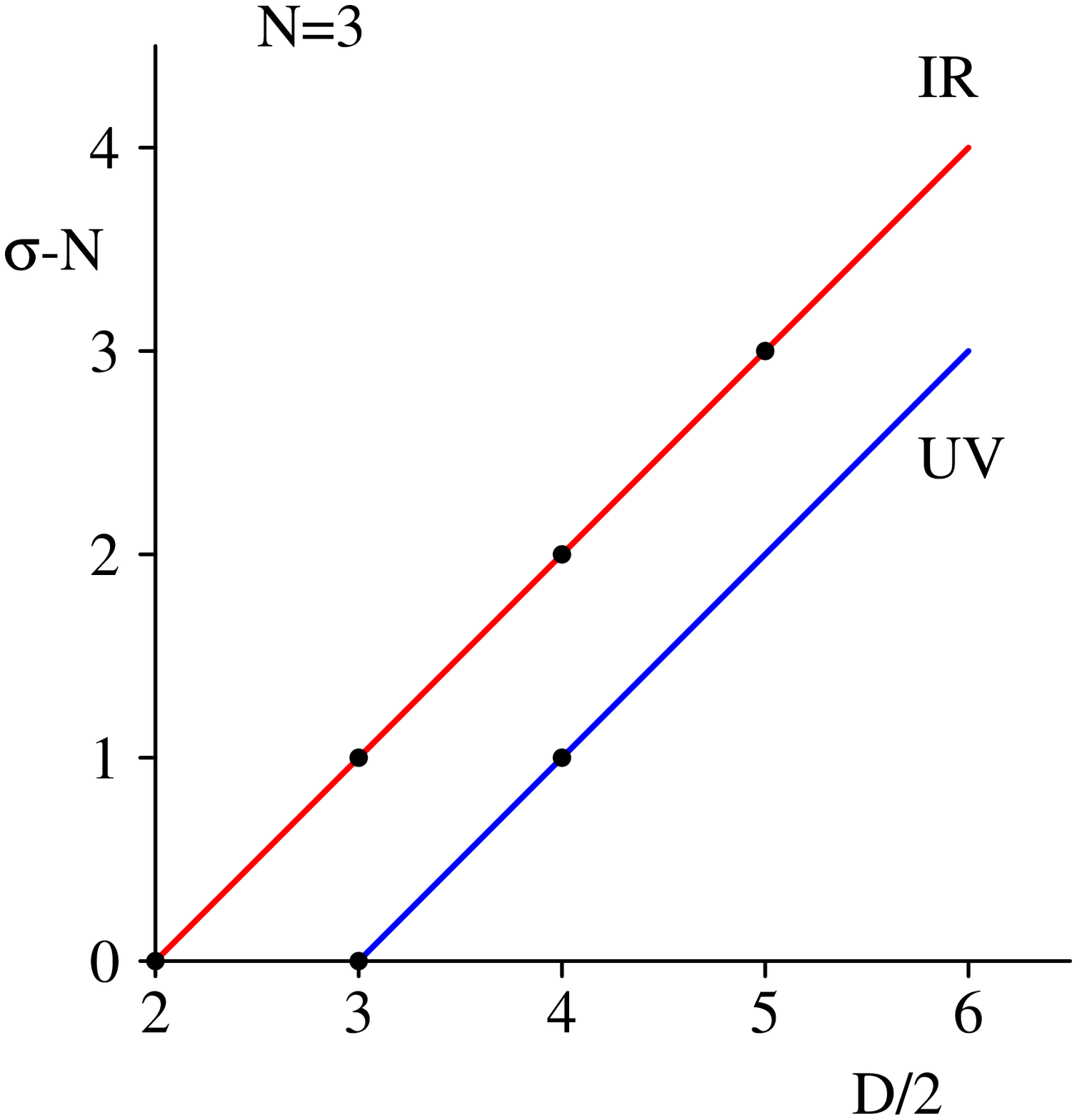}}
\newline
   (c)
\scalebox{.3}{\includegraphics{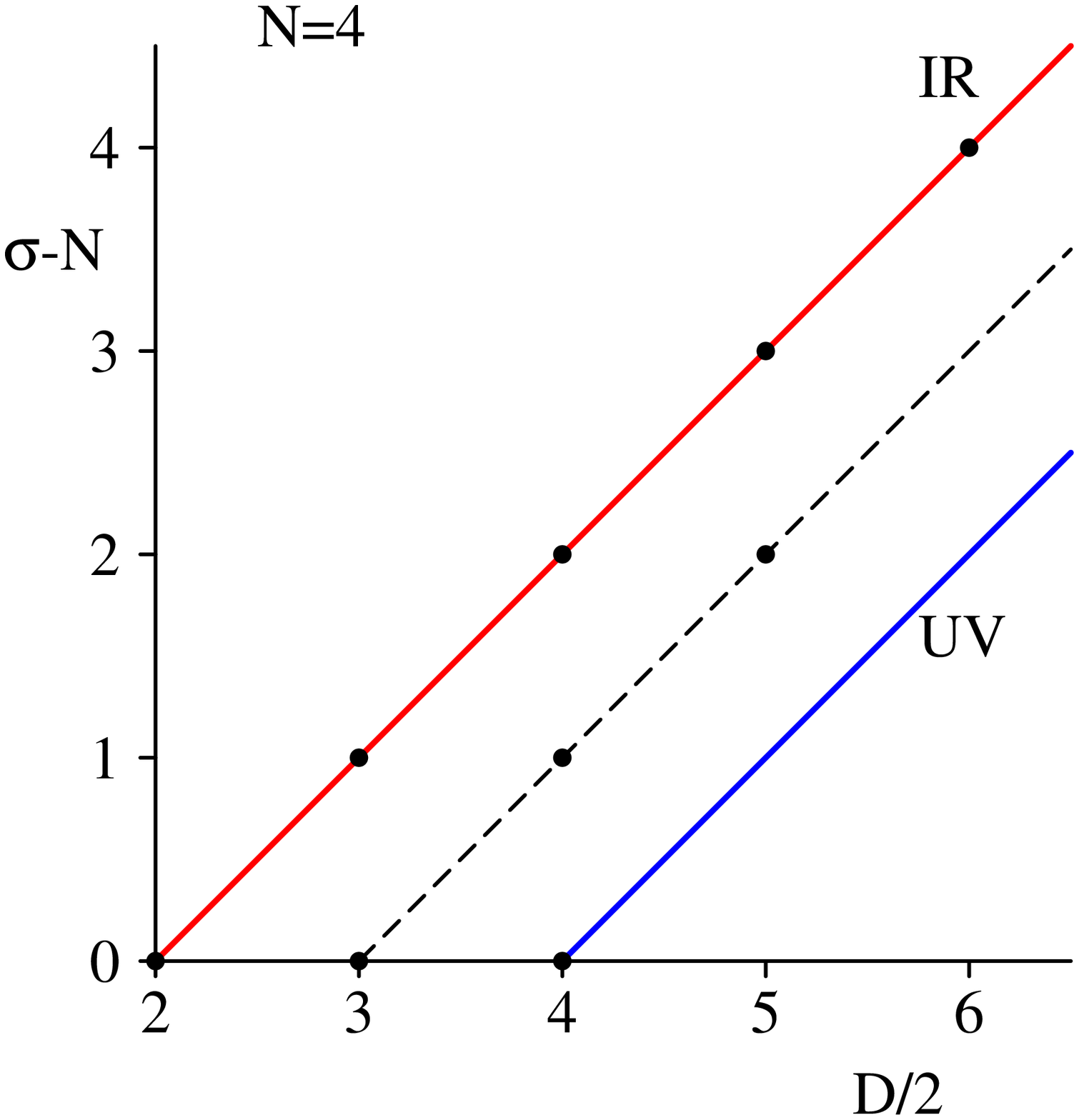}}
   (d)
\scalebox{.3}{\includegraphics{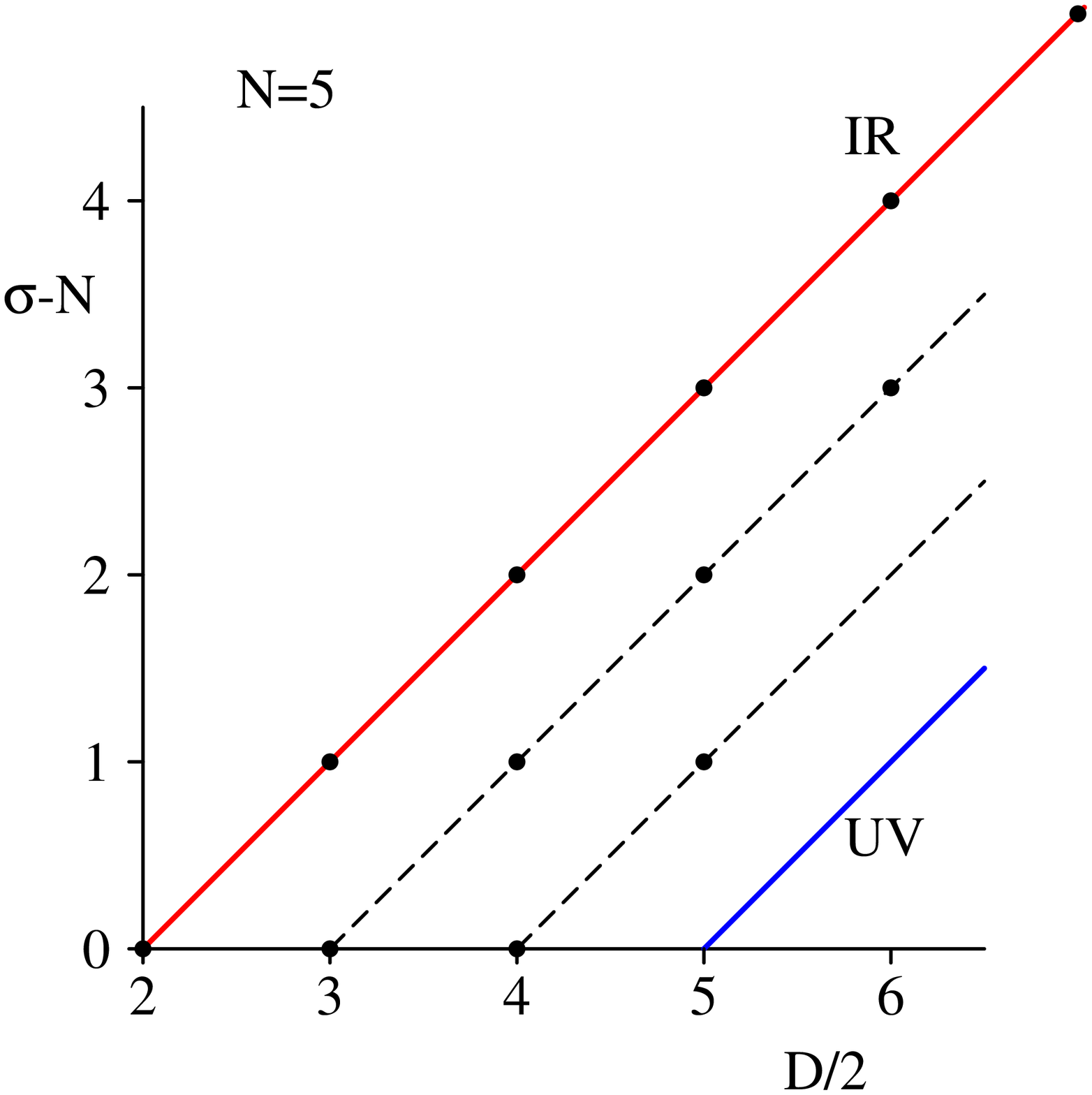}}
\caption{A plot in $(D/2,\sigma)$ for $N=2,3,4,5$.  
The dots represent integrals that
can occur in a Feynman diagram evaluation. The ultraviolet divergent integrals
lie along the lower solid line, while the infrared divergences lie along the
upper solid line. For $N=2$, the IR and UV divergences coincide and are shown
as a magenta line.  Finite integrals lie on the dashed lines.  The ultraviolet
divergent six-dimensional bubble graph (produced by the second rank tensor
bubble) is denoted by a blue circle-cross.}
}

First consider how the singularity conditions of Eqs.~(\ref{eq:IRcond}), (\ref{eq:UVcond}) and
(\ref{eq:fin}) work when $\sigma$ and $D$ vary for
fixed $N$.  This is illustrated in Fig.~\ref{fig:Dsigma} for $N=2,3,4,5$.
In each case, the IR divergent condition of Eq.~(\ref{eq:IRcond}) is shown as a red
line, while Eq.~(\ref{eq:UVcond}) is shown as a blue line.   For bubble graphs
(Fig.~\ref{fig:Dsigma}(a)), the IR and UV lines coincide and are drawn in magenta.  The dots represent
the integrals that appear when Eq.~(\ref{eq:davydychev}) is applied to tensor integrals up to rank $N$.
In all cases, the IR line passes through origin, $\sigma = N$ and $D = 4$,
indicating that the $D=4$ scalar integral may be IR divergent.
We also see that the allowed triangle integrals are always either IR divergent (unless the legs are all
off-shell) or UV divergent.   On the other hand, for $N \geq 4$, families of
finite integrals appear starting with the scalar integral in $D=6$.
\FIGURE[t!]{
\label{fig:Nsigma}
   (a)
\scalebox{.3}{\includegraphics{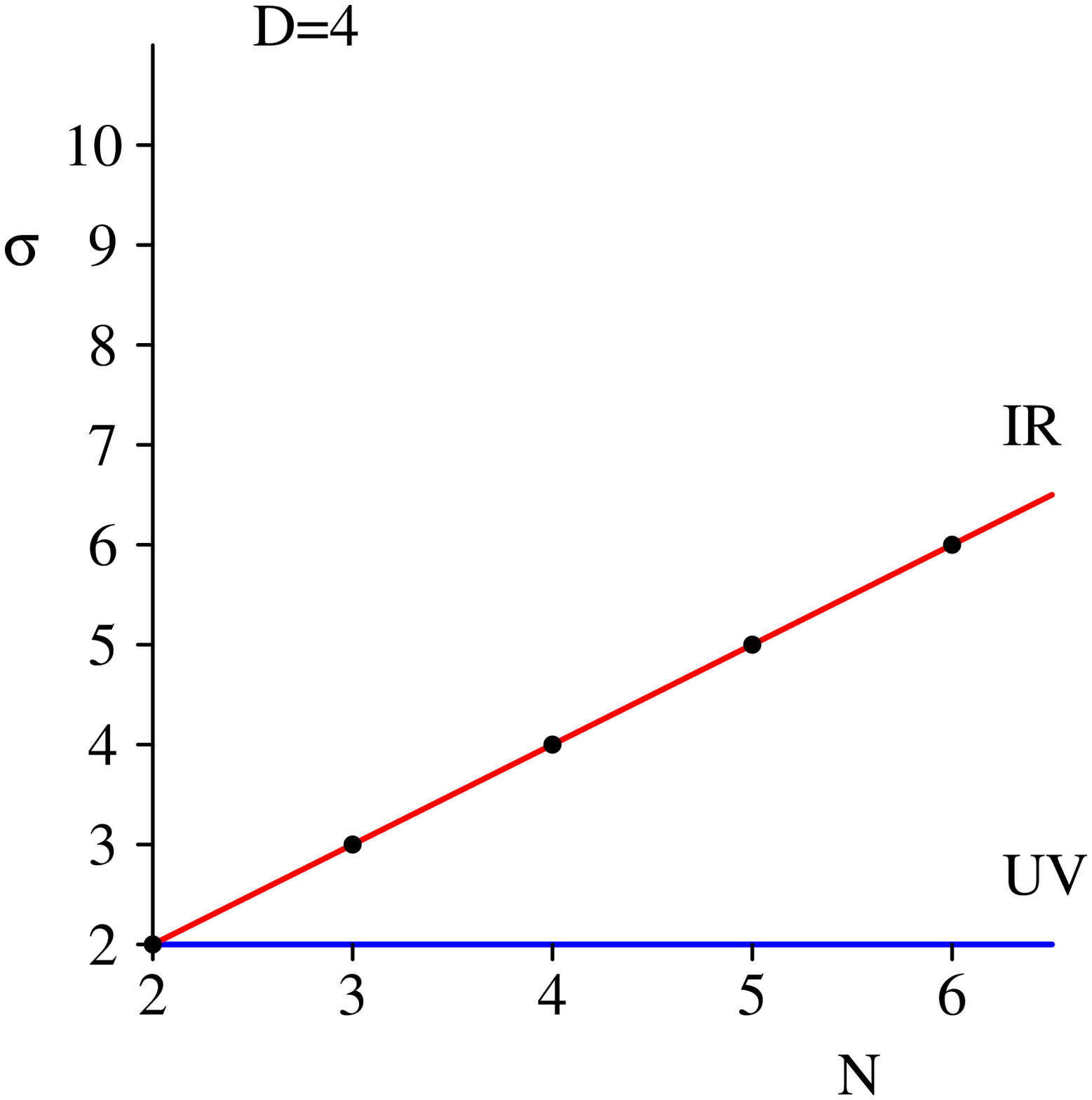}}
   (b)
\scalebox{.3}{\includegraphics{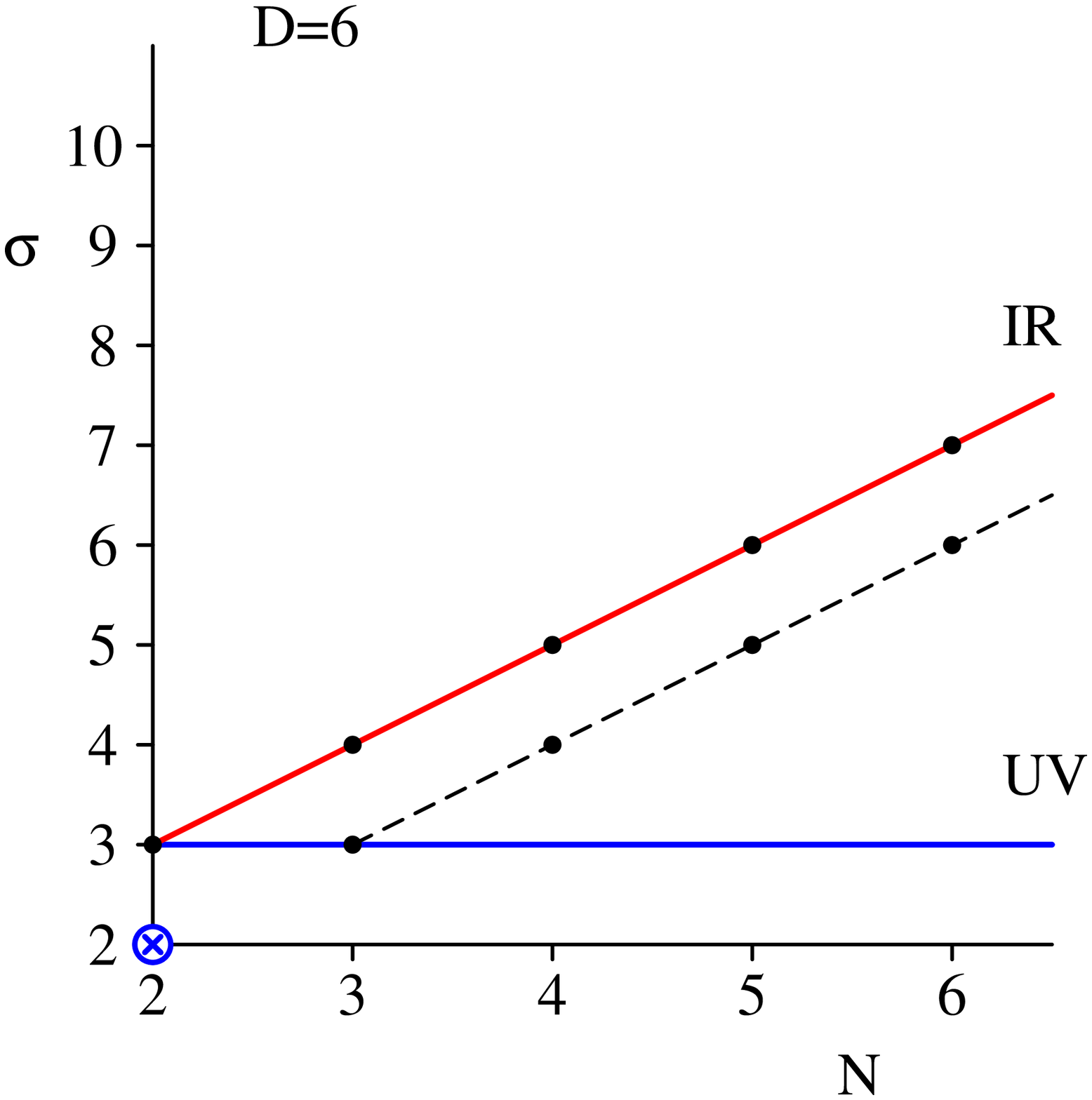}}
\newline
   (c)
\scalebox{.3}{\includegraphics{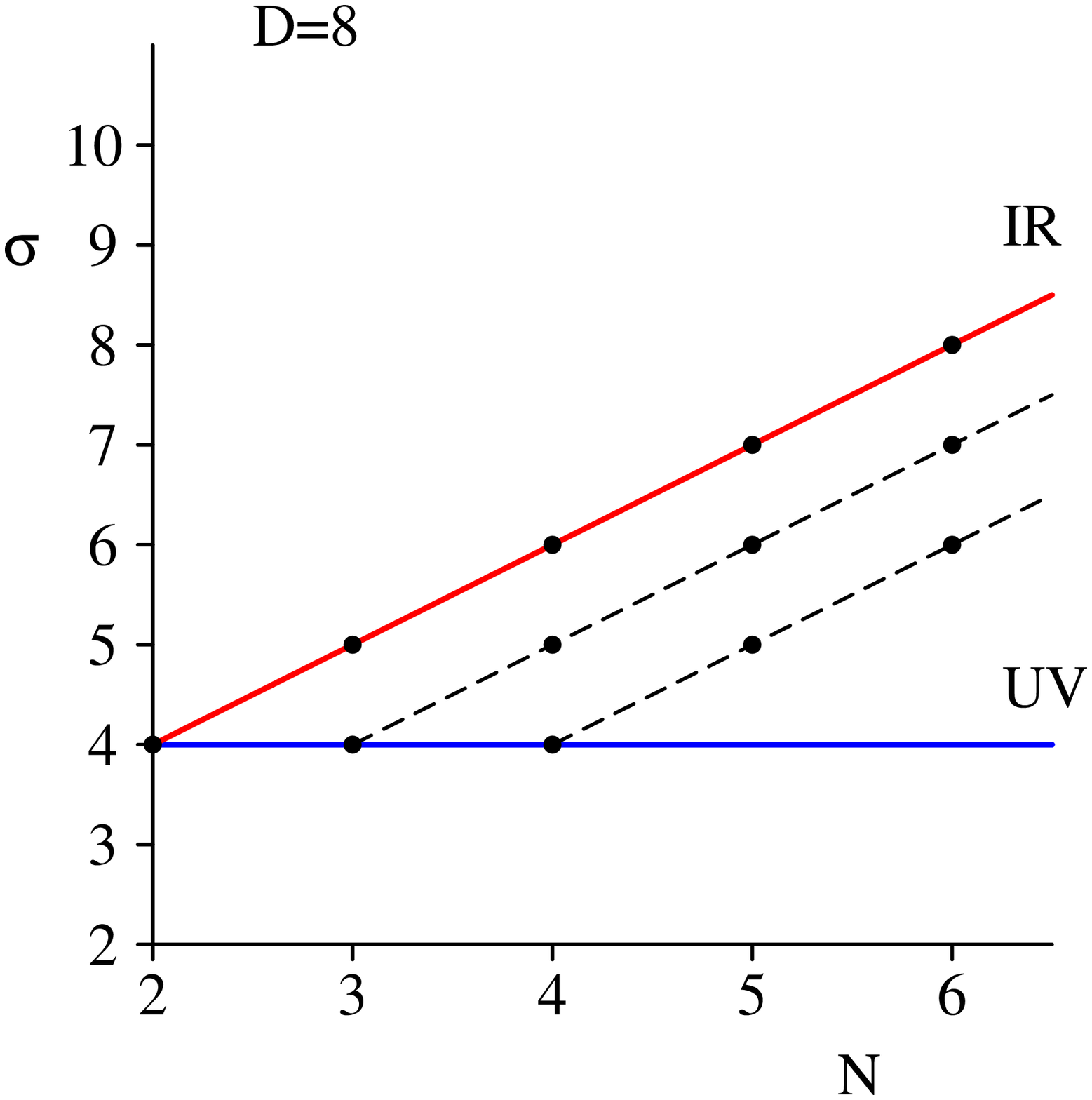}}
   (d)
\scalebox{.3}{\includegraphics{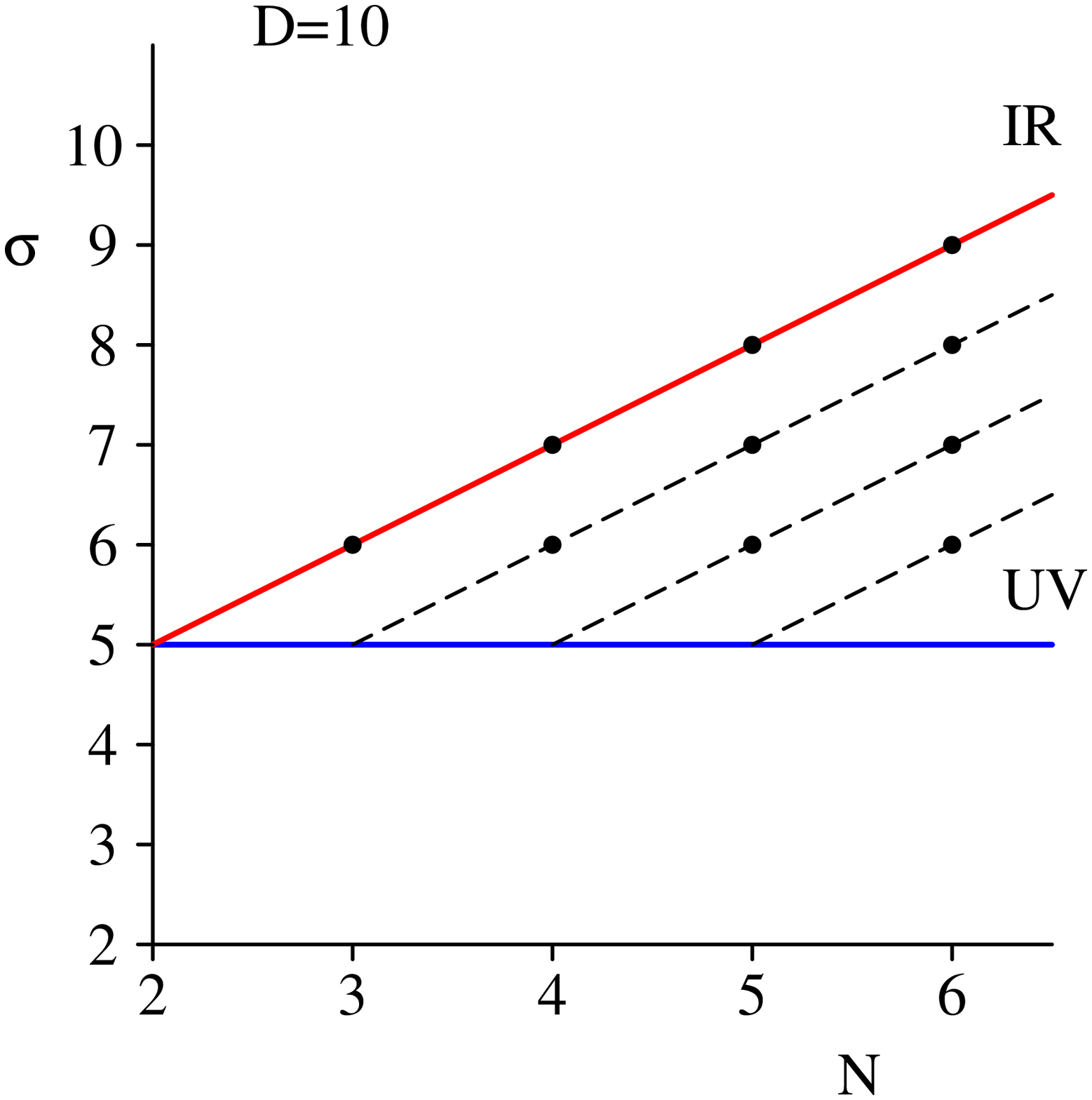}}
\caption{A plot in $(N,\sigma)$ for $D=4,6,8,10$.  The dots represent
integrals that can occur in a Feynman diagram evaluation. The ultraviolet
divergent integrals lie along the blue horizontal line, while the infrared
divergences lie along the red line.   Finite integrals lie on the dashed
lines. The ultraviolet divergent six-dimensional bubble graph 
(produced by the second rank tensor
bubble) is denoted by a blue circle-cross. }
}

It is also instructive to consider the finiteness of integrals of varying $N$
with fixed dimension.   This is shown in Fig.~\ref{fig:Nsigma} where the IR and
UV conditions are again shown red and blue while the dots correspond to 
integrals produced by Eq.~(\ref{eq:davydychev}). 
Fig.~\ref{fig:Nsigma}(a) shows that all four-dimensional integrals
can be divergent while the first integrals that are guaranteed to be
finite independent of whether the legs are on-shell or not are higher dimension
box graphs.

Reading off the allowed UV divergent integrals from Fig.~\ref{fig:Nsigma},
we see that a suitable (but overcomplete) basis set is
\begin{eqnarray}\label{eq:UVset}
{\cal I}^{UV}&=&\left\{I_4(D=8-2\e;1,1,1,1),I_3(D=8-2\e;2,1,1),I_2(D=8-2\e;3,1),
\right.\nonumber\\ && \left.
I_2(D=8-2\e;2,2),I_3(D=6-2\e;1,1,1),I_2(D=6-2\e;2,1),
\right.\nonumber\\ && \left.
I_2(D=6-2\e;1,1),I_2(D=4-2\e;1,1)\right\}
\end{eqnarray}
such that
\begin{equation}\label{eq:UVcoef}
{\cal A}_M(p_1,p_2,\ldots,p_M)=
\sum_i K_i^{IR}{\cal I}^{IR}_i(\{q_j\})
+\sum_i K_i^{fin}{\cal I}^{fin}_i(\{q_j\})
+\sum_{i=1}^8 K_i^{UV}{\cal I}_i^{UV}
(\{q_j\}).
\end{equation}
While the UV divergent functions are all known analytically~\cite{Abox,BD,Ndim}, 
the basis sets of finite and IR divergent integrals have an
infinite number of members.
To derive finite basis sets of known integrals for the latter two classes
we employ recursion relations.

\subsection{Recursion Relations}
\label{sec:recursion}

In this subsection we show how to map the higher dimensional finite and IR
divergent integrals onto a basis set of known finite and divergent
integrals. This will fully specify Eq.~(\ref{eq:ansatz}) and produce the
explicit set of master integrals $\{{\cal I}^{UV},{\cal I}^{IR},{\cal I}^{fin}\}$.

Because the external momenta are 4-dimensional, the recursion relations 
come in two distinct classes that depend  on the number of external legs.
When the number of legs is five or less, the external momenta
form a vector basis for the resulting tensor structure. However, for
more than five external lines, the set of external vectors is over-complete
and the net result is that the recursion relations change for $N \geq 6$.

First, we introduce some notation and 
define the symmetric kinematic matrix 
\begin{equation}
S_{ij}=(q_i-q_j)^2.
\end{equation}
By adding a row (or
column) over the inverse of the kinematic matrix we make the further
definitions (provided the inverse of the kinematic matrix exists),
\begin{equation}
b_i=\sum_j S^{-1}_{ij},\qquad\qquad
B=\sum_i b_i=\sum_{ij} S^{-1}_{ij}.
\end{equation}
The Gram matrix
\begin{equation}
G_{ij}=2 q_i\cdot q_j,
\end{equation}
is closely related to the kinematic matrix.

The recursion relations needed to relate all possible integrals to the basis
sets are formulated as follows:\\
%\begin{itemize}
%\item[I.]
 
\noindent{{\bf I. $N\leq 5$}}: $\det(S)\neq 0$, $\det(G)\neq 0$ \newline

We will use four recursion relations in this case, one basic recursion relation 
and three derived, composite recursion relations.  

The basic recursion relation is obtained by integration by parts as explained
in appendix~\ref{app:recursion}. It is given by,
\begin{eqnarray}
\label{eq:recursion1}
\lefteqn{(\nu_k-1) I_N(D;\{\nu_l\})}\nonumber \\
&=&-\sum_{i=1}^N S_{ki}^{-1} I_N(D-2;\{\nu_l-\delta_{li}-\delta_{lk}\})
-b_k\left(D-\sigma\right)I_N(D;\{\nu_l-\delta_{lk}\}),\nonumber \\
\end{eqnarray}
where $\sigma$ is defined in Eq.~(\ref{eq:sigmadef}).
This relation either
reduces both the dimension $D$ and the accompanying
value of $\sigma$ by two, 
or it keeps $D$ fixed and reduces $\sigma$ by one unit.
Its action is indicated by the red lines in Fig.~\ref{fig:path}. 
This identity is vital for extracting the infrared singularities from the
integrals produced by the Davydychev decomposition of 
Eq.~(\ref{eq:davydychev}).

Note that for the case $N=3$ in Fig.~\ref{fig:path}(a) we assume that 
$\det(S)\neq 0$ which is only the case if all three external momenta
are off-shell. If any of the external momenta
is an on-shell lightcone vector, the inverse of the kinematic matrix does
not exist and Eq.~(\ref{eq:recursion1}) is invalid. For practical purposes,
this is not a problem because analytic expressions for arbitrary dimension
and powers of propagators exist for triangle integrals with at least one
on-shell leg and massless internal propagators (see Appendix~\ref{app:analytic}).  
In four dimensions, the triangle integral with three off-shell legs and unit propagators
is finite 
and this serves as a reminder
that integrals on the ``IR'' lines in Fig.~\ref{fig:path} are not
necessarily IR singular, rather they can be IR singular if some of
the external momenta are lightlike.

Eq.~(\ref{eq:recursion1}) is a more detailed equation than usually quoted
in the literature and it can only be applied to scalar integrals with
at least one index $\nu_k>1$. However, we can use it to
derive the standard recursion relation~\cite{binoth},
\begin{equation}
\label{eq:recursion2}
\left(D-1-\sigma\right)\,B\,I_N(D;\{\nu_l\})=I_N(D-2;\{\nu_l\})
-\sum_{i=1}^Nb_iI_N(D-2;\{\nu_l-\delta_{li}\}).
\end{equation}
Details of the derivation can be found in Appendix~\ref{app:recursion}.
Eq.~(\ref{eq:recursion2}) reduces $D$ by 2 and $\sigma$ by 1.  
Its action is illustrated by the blue lines shown in Fig.~\ref{fig:path}. 
For the case that all $\nu_l=1$ and $D=4-2\e$, 
we will use this recursion relation to extract the infrared singularities
from graphs with at least one on-shell leg
as pinched integrals together with scalar integrals in $D=6-2\e$ dimensions,
\begin{equation}
\label{eq:recursion2a}
I_N(D;\{\nu_l\})=\left(D+1-\sigma\right)\,B\,I_N(D+2;\{\nu_l\})
+\sum_{i=1}^Nb_iI_N(D;\{\nu_l-\delta_{li}\}).
\end{equation}
Its action is illustrated by the magenta line in Fig.~\ref{fig:path}. 

\FIGURE[t!]{
\label{fig:path}
   (a)
\scalebox{.3}{\includegraphics{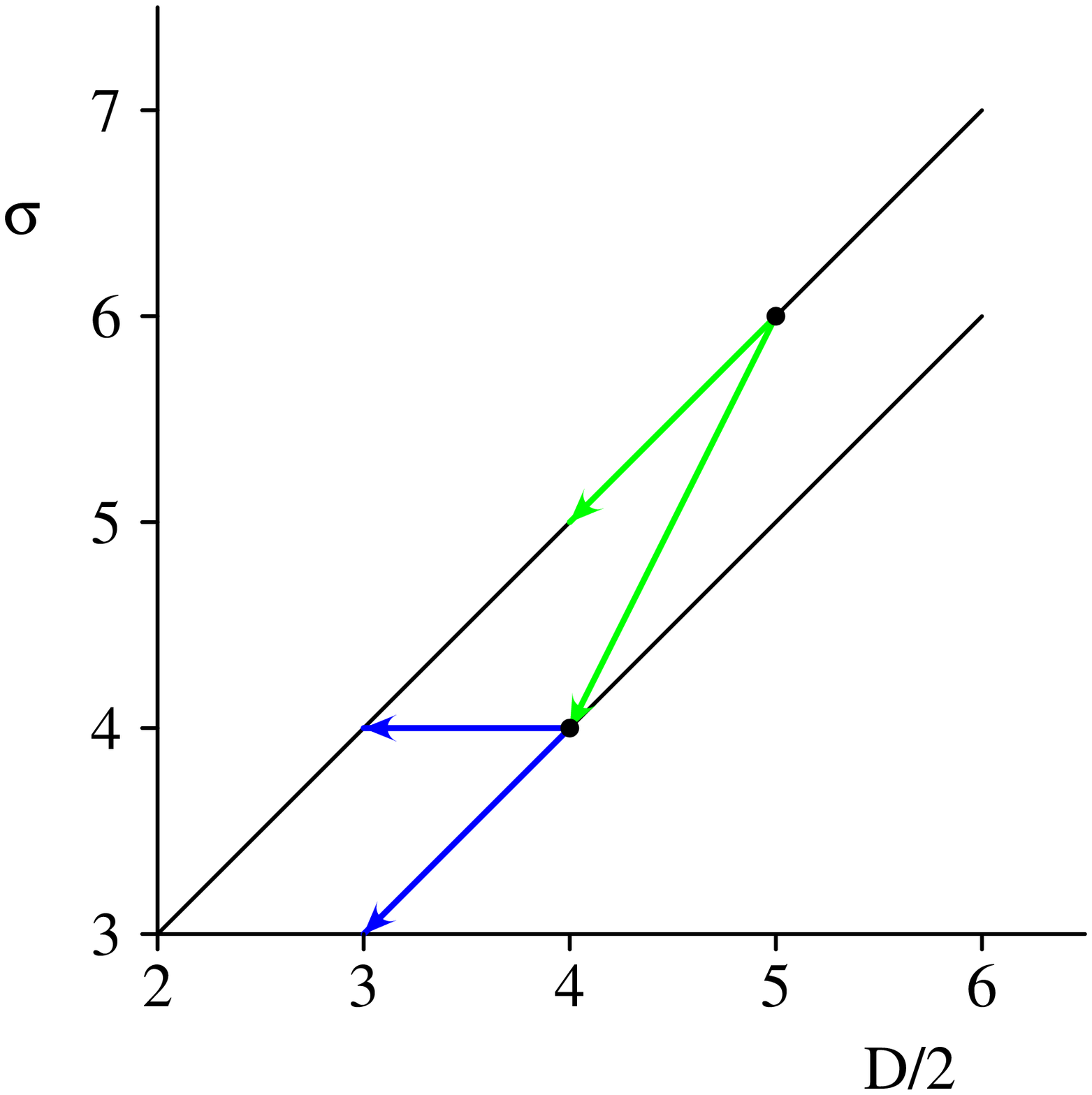}}
   (b)
\scalebox{.3}{\includegraphics{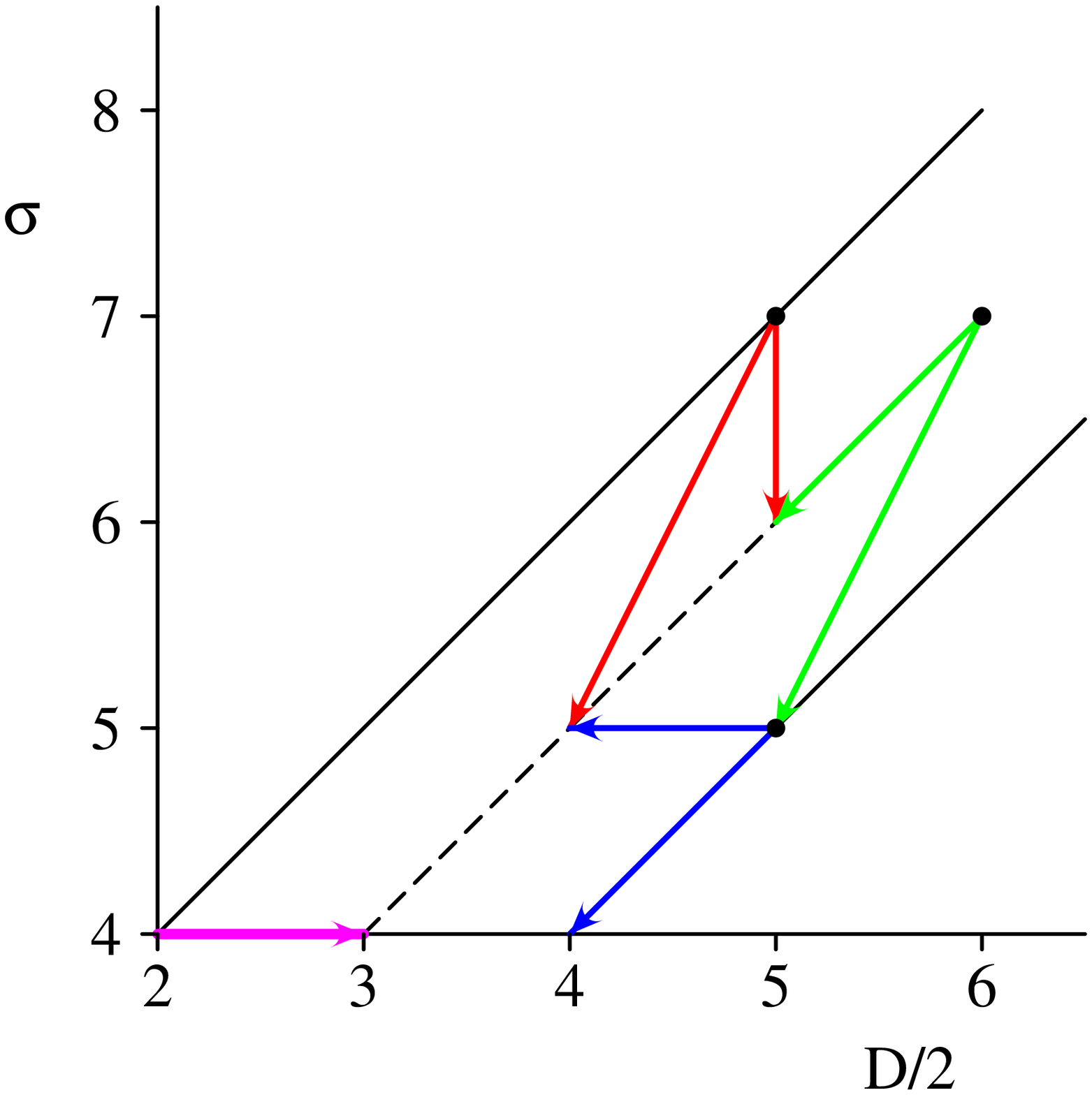}}
\newline
   (c)
\scalebox{.3}{\includegraphics{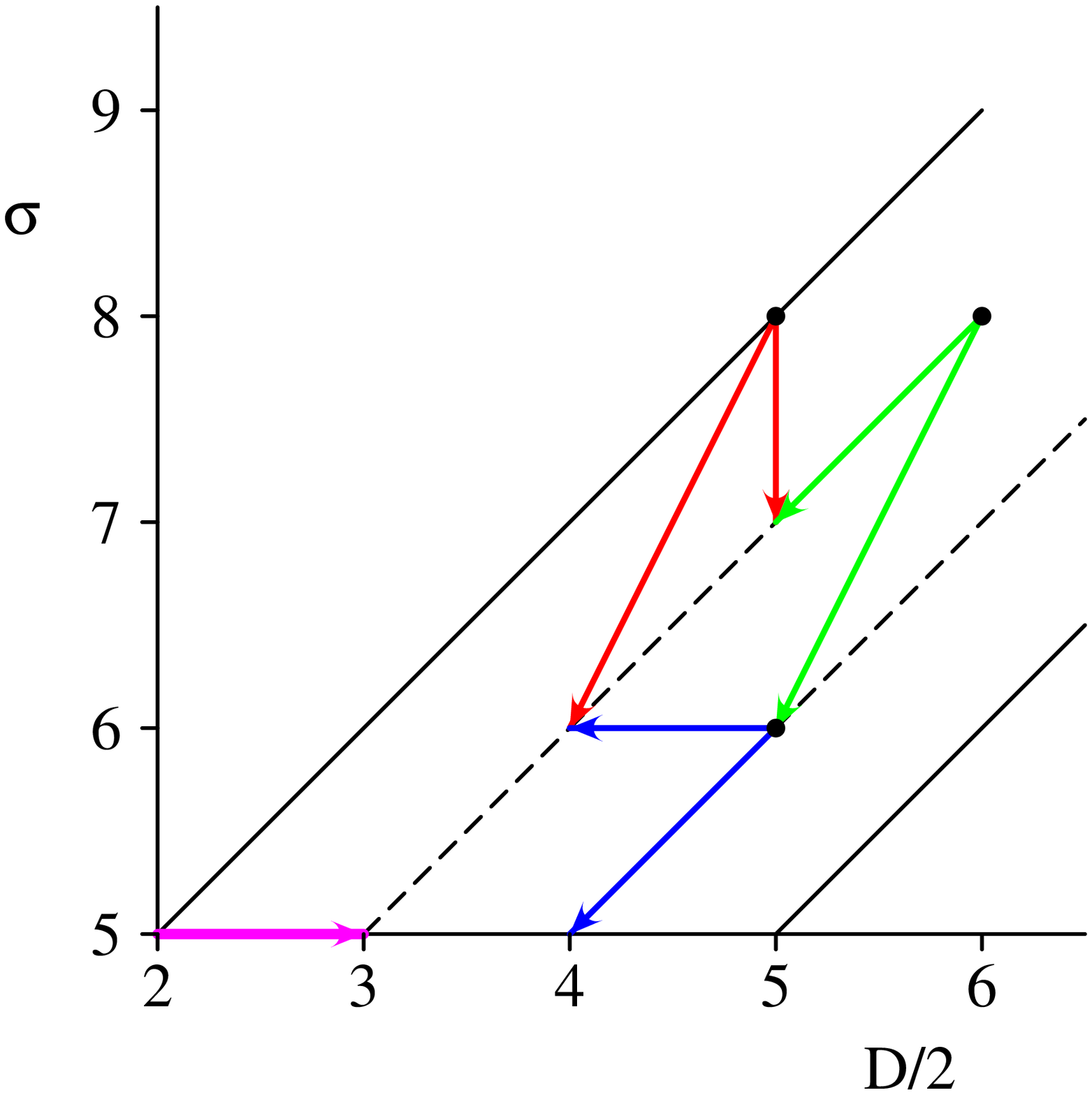}}
\caption{A plot in $(D/2,\sigma)$ for (a) the finite triangle graph, $N=3$,
(b) the box graph, $N=4$ and (c) the
pentagon graph, $N=5$.   The ultraviolet
divergent integrals lie along the lower solid line, while for $N > 3$
any infrared
divergences lie along the upper solid line.   Finite integrals lie on the dashed
lines.  The action of
Eqs.~(\ref{eq:recursion1}),~(\ref{eq:recursion2})~(\ref{eq:recursion2a}) 
and~(\ref{eq:recursion3}) 
are shown red, blue, magenta and green
respectively.}
}

By combining Eqs.~(\ref{eq:recursion1}) and (\ref{eq:recursion2}) 
we can eliminate
the dimensional prefactor to yield the fourth recursion relation,
\begin{eqnarray}
\label{eq:recursion3}
\lefteqn{(\nu_k-1) I_N(D;\{\nu_l\})}\nonumber \\
&=&-\frac{b_k}{B}I_N(D-2;\{\nu_l-\delta_{lk}\})
+\sum_{i=1}^N\left(\frac{b_kb_i}{B}-S_{ki}^{-1}\right) 
I_N(D-2;\{\nu_l-\delta_{li}-\delta_{lk}\}).\nonumber \\
\end{eqnarray}
This identity reduces $D$ by 2 and $\sigma$ by either 1 or 2 units and
its action is indicated in Fig.~\ref{fig:path} by the green lines.

When $N=5$ it is easy to avoid the UV line because of the presence of two
types of finite integrals ($D/2=\sigma-1$ and $D/2=\sigma-2$). 
However, this is not the case for $N=3$ and $N=4$ and we
are forced to use Eq.~(\ref{eq:recursion3}) which appears to
produce UV divergent integrals
from finite integrals. This implies that in the sum over the produced
UV divergent integrals the divergences must cancel. Indeed, because 
$$\sum_{i=1}^N\left(\frac{b_kb_i}{B}-S^{-1}_{ki}\right)=0$$ these
artificial UV divergences in the individual integrals cancel
in the sum as expected. This means we 
replace each UV divergent integral in Eq.~(\ref{eq:recursion3})
by a UV finite regulated integral,
\begin{equation}
\tilde{I}_N^{UV}(D=2\sigma;\{\nu_l\})=I_N^{UV}(D=2\sigma;\{\nu_l\})-R(D,\sigma),
\end{equation}
where the scaleless regulator term depends on
the dimension and $\sigma$ with the constraint $D=2\sigma$,
\begin{equation}
R(D,\sigma)=\frac{1}{\e}\frac{(-1)^{\sigma}}{\Gamma(D-\sigma)}.
\end{equation}
After the subtraction the integral is rendered finite.
Of course,  for this replacement 
to be useful we must require that
Eq.~(\ref{eq:recursion2}) is still valid for the 
regulated integrals $\tilde{I}_N$.
Therefore, the regulator function {\it must} obey 
\begin{equation}
\left(D-1-\sigma\right)\,B\,R(D,\sigma)=-\sum_{i=1}^Nb_iR(D-2,\sigma),
\end{equation}
or equivalently
\begin{equation}
\left(D-\sigma-1\right) R(D,\sigma)= -R(D-2,\sigma),
\end{equation}
which is trivially verified.

The application of the various recursion relations is clear from
the figures:
\begin{itemize}
\item If $N\neq 3$ and $D=4$ apply Eq.~(\ref{eq:recursion2a}) (magenta arrow)
\item If $N\neq 3$ and integral is on the IR line ($D/2=\sigma-N+2$) 
apply Eq.~(\ref{eq:recursion1}) (red arrows)
\item In the other cases one can apply  
Eqs.~(\ref{eq:recursion2}) (blue arrows) and (\ref{eq:recursion3}) (green arrows) with
the restriction $\sigma-N+2<D/2\leq\sigma$. Any ``fake'' UV divergent integrals 
generated can be regulated as explained in the previous paragraph.
\end{itemize}
These simple rules are sufficient to reduce all integrals 
with $N \leq 5$ 
and non-exceptional kinematics to the basis set of integrals.\\

\noindent{{\bf II. $N=6$}}: $\det(S)\neq 0$, $\det(G)=0$ \newline

For $N=6$ the kinematic matrix $S_{ij}$ is still invertible. This
means that the same recursion relations hold as in the $N<6$ case.
However,
because $\det(G)=0$, $B=\sum_ib_i=0$~\cite{binoth}.
This does not affect the basic recursion relation 
given in Eq.~(\ref{eq:recursion1}), but does alter
the composite recursion relations and
Eqs.~(\ref{eq:recursion2})--(\ref{eq:recursion3})
all collapse to yield
\begin{equation}
\label{eq:recursion4}
I_6(D;\{\nu_l\})=\sum_{i=1}^6b_iI_6(D;\{\nu_l-\delta_{li}\})\ .
\end{equation}
This equation preserves the value of $D$, but reduces $\sigma$ by unity.
Its action is indicated by the green arrow in Fig.~\ref{fig:path6}.
An additional equation that reduces $D$ is also required and can be 
obtained by combining
Eqs.~(\ref{eq:recursion1}) and (\ref{eq:recursion4}),
\begin{eqnarray}
\label{eq:recursion4a}
\lefteqn{b_k(D-\sigma+\nu_k-1)\,I_6(D;\{\nu_l\})=} \nonumber \\ &&
-\nu_k\sum_{i\neq k}^6 b_i I_6(D;\{\nu_l+\delta_{lk}-\delta_{li}\})
-\sum_{i=1}^6 S_{ki}^{-1} I_6(D-2;\{\nu_l-\delta_{li}\}).
\end{eqnarray}
This recursion relation works in 2 different manners. 
The second term reduces $D$ by 2 and $\sigma$ by unity and acts in a similar
manner to the identities for $N\leq 5$ case and is illustrated
by the magenta arrow in Fig.~\ref{fig:path6}.
On the other hand, the first term preserves both $D$ and $\sigma$
while reducing $N$.    

The application of the recursion relations for this case are straightforward.
\begin{itemize}
\item If $N=\sigma$ apply Eq.~(\ref{eq:recursion4}).
\item If $D/2=\sigma-N+2$ (i.e. on the IR line) 
apply Eq.~(\ref{eq:recursion1}) (red arrow).
\item If $D/2<\sigma-1$ 
apply Eq.~(\ref{eq:recursion4}) or (\ref{eq:recursion1})
 (green or red arrow).
\item If $D/2=\sigma-1$ apply Eq.~(\ref{eq:recursion4a}) (magenta line). The
 index $k$ is chosen such that $\nu_k=\max_{l}(\nu_l)$ (this is not necessarily a unique
choice).
\end{itemize}

These simple rules are sufficient to reduce all integrals 
with $N=6$ and non-exceptional kinematics to $N=5$ integrals.\\

\FIGURE[t!]{
\label{fig:path6}
\scalebox{.5}{\includegraphics{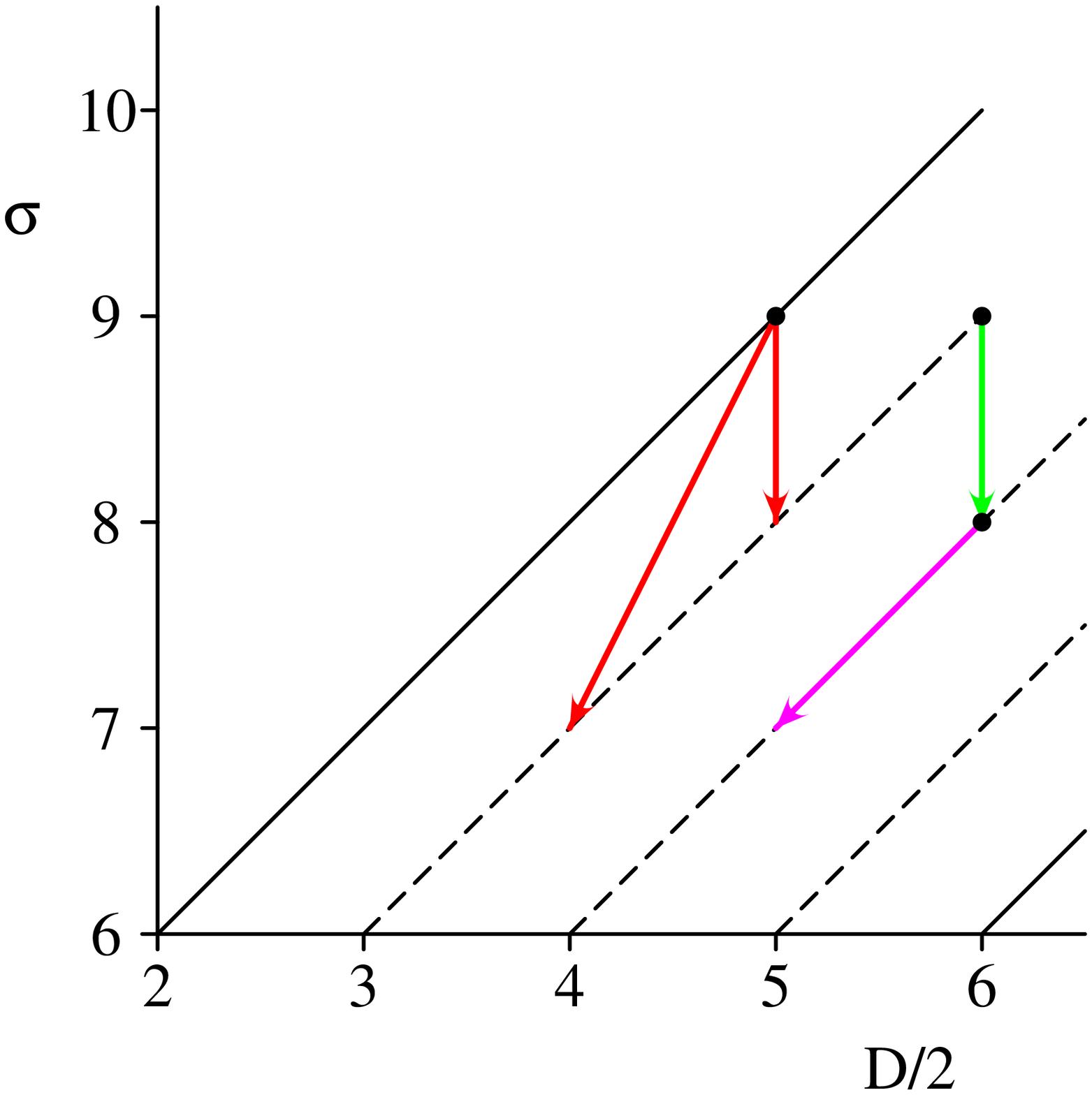}}
\caption{A plot in $(D/2,\sigma)$ for hexagon graphs, $N=6$.   
The ultraviolet
divergent integrals lie along the lower solid line, while the infrared
divergences lie along the upper solid line.   
Finite integrals lie on the dashed
lines.  The action of
Eqs.~(\ref{eq:recursion1}),~(\ref{eq:recursion4})  
and~(\ref{eq:recursion4a}) 
are shown red, green and magenta
respectively. Note that integrals produced by pinching 
so that $N < 6$ are not shown.}
}

\noindent{{\bf III. $N\geq 7$}}: $\det(S)=0$, $\det(G)=0$ \newline

In this case, the inverse of the kinematic matrix 
does not exist.
This invalidates the use of 
Eqs.~(\ref{eq:recursion1})--(\ref{eq:recursion3}) for
$N\geq 7$. 

However, the fact that the kinematic matrix is made up of 4-dimensional
vectors which obey momentum conservation leads to additional relations.
To derive these new types of recursion relations valid for
these high values of $N$, we use a generalized inverse based on the
singular value decomposition of the Gram matrix.

The first basic recursion relation is based on the kernel of $S$. 
By solving $(S.z)_i=0$ for $z$ we obtain~\cite{Nizic} 
\begin{equation}
\label{eq:recursion5}
z_k\,I_N(D;\{\nu_l\})=-\sum_{i\neq k} z_i I_N(D;\{\nu_l+\delta_{lk}-\delta_{li}\}),
\end{equation}
or alternatively,
\begin{equation}
\label{eq:recursion5a}
\sum_i z_i I_N(D;\{\nu_l+\delta_{lk}-\delta_{li}\})=0,
\end{equation}
where the explicit $z_i$ are given in Appendix \ref{app:recursion}. 
An important property is that $\sum_i z_i=0$.  
Eq.~(\ref{eq:recursion5}) keeps
$D$ and $\sigma$ constant, but can pinch out particular propagators to reduce
$N$.   Its effect is illustrated by the magenta arrows in Fig.~\ref{fig:path7}.

The existence of the second recursion relation  is highly non-trivial. It is
based on the fact that the unit vector $I_i=(1,1,\ldots,1)$ 
lies in the range of the singular
matrix $S$, i.e. the equation $(S.r)_i=I_i$ has a solution for $r_i$. The
existence of the solution and the explicit construction of the vector $r_i$ are
detailed in Appendix~\ref{app:recursion} and leads to~\cite{Nizic}
\begin{equation}
\label{eq:recursion6}
I_N(D;\{\nu_l\})=\sum_i r_i I_N(D;\{\nu_l-\delta_{li}\})
\end{equation}
where $\sum_i r_i=0$. 
While in functional form this recursion relation is equivalent to
Eq.~(\ref{eq:recursion4}), its origin is rather different. This is reflected in
the fact that the unique coefficients $b_i=\sum_j S_{ij}^{-1}$ are unrelated
to the many possible and equivalent sets of $r_i$ coefficients.
For example,  replacing $r_i\rightarrow r_i+\alpha z_i$ leaves 
Eq.~(\ref{eq:recursion6}) unaltered. 
Eq.~(\ref{eq:recursion6}) also preserves $D$, but reduces $\sigma$ (and possibly $N$) by unity as illustrated by the cyan lines in Fig.~\ref{fig:path7}.

\FIGURE[t!]{
\label{fig:path7}
\scalebox{.5}{\includegraphics{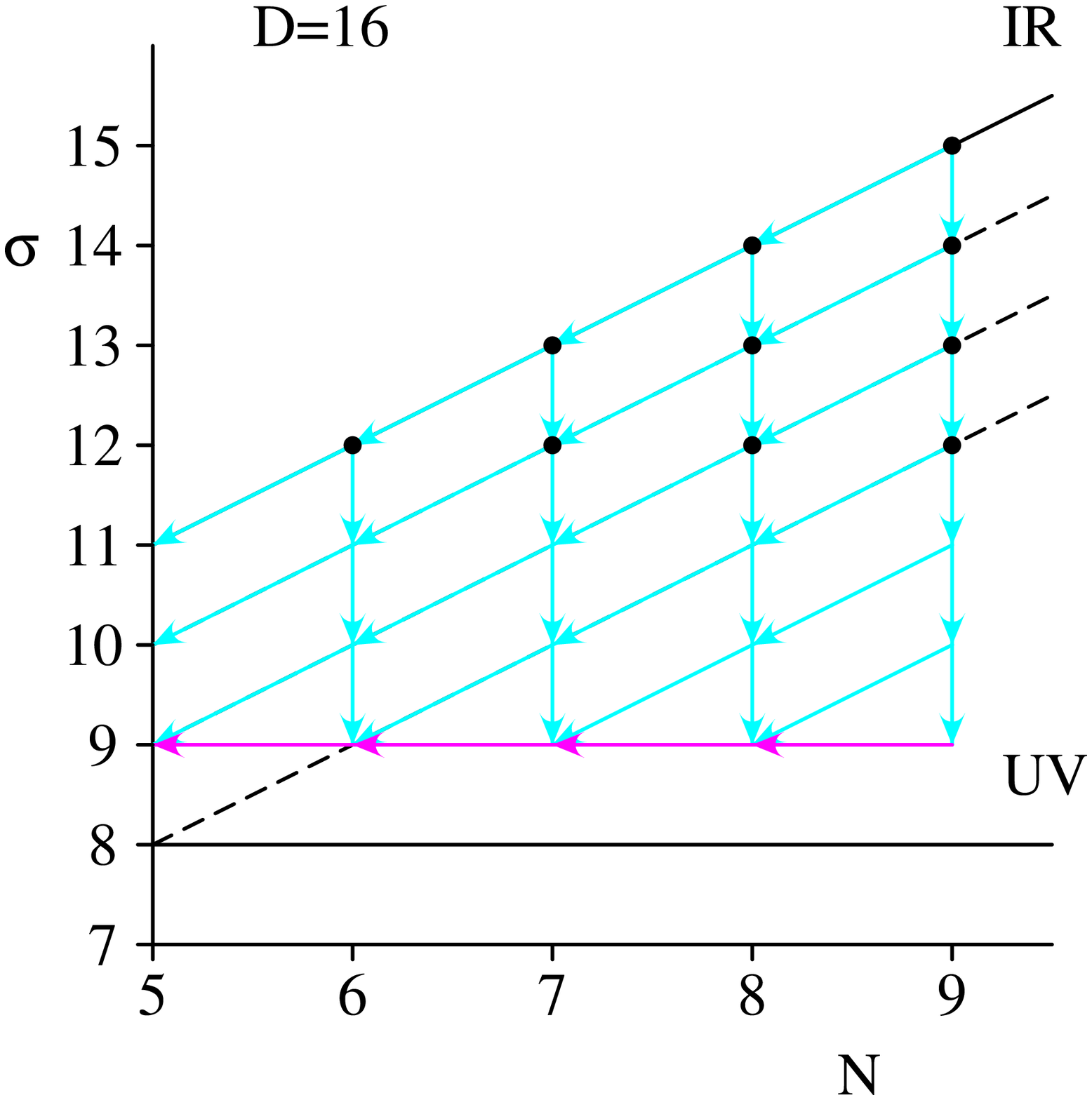}}
\caption{A plot in $(N,\sigma)$ for $D=16$ illustrating the procedure for
evaluating integrals with $N \geq 6$.  The dots represent the integrals 
produced  
in the Davydychev decomposition of rank 6, 7, 8 and 9 tensor integrals.
 The ultraviolet divergent integrals lie along the lower solid blue 
line, while the infrared
divergences lie along the upper solid red line.  The action of 
Eq.~(\ref{eq:recursion5}) and
Eq.~(\ref{eq:recursion6}) are shown
as  magenta and cyan respectively.}
}

The application of the recursion relations for this case 
are similar to the $N=6$ case:

\begin{itemize}
\item If $N=\sigma$ use Eq.~(\ref{eq:recursion6})
\item If $D/2<\sigma-1$ apply Eq.~(\ref{eq:recursion6})
 (cyan arrows).
\item If $D/2=\sigma-1$ apply Eq.~(\ref{eq:recursion5}) (magenta line). The
 index $k$ is chosen such that $\nu_k=\max_{l}(\nu_l)$ (this is not necessarily a unique
choice).
\end{itemize}

These simple rules are sufficient to reduce all integrals 
with $N\geq 7$ and non-exceptional kinematics to $N=6$ integrals.

\subsection{The basis set of integrals}
\label{sec:basis}

The net result of applying the recursion relations is that any 
amplitude can be written as,
\begin{eqnarray} \label{eq:genamp}
{\cal A}_M(p_1,p_2,\ldots,p_M)&=&
\sum_{\nu_1\nu_2\nu_3} K_{\nu_1\nu_2\nu_3}^{IR} I_3^{IR}(D=2(\sigma-1);\nu_1,\nu_2,\nu_3)\nonumber \\
&+&\sum_{\{\nu_\ell\}} K_{\{\nu_\ell\}}^{fin}
\tilde{I}_N^{UV}(D=2\sigma;\{\nu_\ell\})
\nonumber\\
&+&\sum_{\rm triangles} K_3^{fin} I_3^{fin}(D=4;1,1,1)\nonumber \\
&+&\sum_{\rm boxes} K_4^{fin} I_4^{fin}(D=6;1,1,1,1)\nonumber \\
&+&\sum_{\rm pentagons} K_5^{fin} I_5^{fin}(D=6;1,1,1,1,1)\nonumber \\ 
&+&\sum_{i=1}^8 K_i^{UV} {\cal I}_i^{UV}.
\end{eqnarray}
Note that the absence of new basis integrals for $N\geq 6$ is intimately related to
the dimensionality of space-time (i.e. $\det(G)=0$ for $N\geq 6$).
All of the integrals appearing in this basis set are analytically known. 
The IR and UV divergent
integrals with massless internal propagators are known analytically for 
arbitrary powers of the
propagators and arbitrary dimensions~\cite{BD,Ndim}. 
For the sake of completeness, we list them in the Appendix~\ref{app:analytic}.
Note that the regulated UV divergent triangles
$\tilde{I}_3^{UV}$ are produced by the action of Eqs.~(\ref{eq:recursion3}) 
and (\ref{eq:recursion4}) and the sum is therefore
finite.

As can be seen in Fig.~\ref{fig:Dsigma},
UV divergent integrals occur for 2-point,  3-point  
and 4-point functions. By applying  
Eq.~(\ref{eq:recursion2}) to UV divergent 4-point functions,
the UV divergences are moved to 3-point functions. This means the UV basis set
of Eq.~(\ref{eq:UVset}) is reduced to only 2-point and 3-point functions.  
Analytic formulae for these UV divergent integrals are listed in Appendix~\ref{app:analytic}.

The  finite 4-dimensional triangle integral with off-shell legs, $I_3^{fin}$,
and the 6-dimensional box graphs are also known in terms of logarithms and 
dilogarithms~\cite{Abox,BD}.  Furthermore,  it is an empirical
fact that in final expressions for physical quantities, the 
6-dimensional pentagons do not appear~\cite{Abox,binoth,Nizic}. 
This means $K_5^{fin}={\cal O}(\e)$ and we do not need to know 
$I_5(D=6;1,1,1,1,1)$ for NLO calculations.

\section{Determining the Infrared Divergent Contributions}
\label{sec:subtract}

In this section we derive the techniques for determining the coefficients
$K^{IR}$ that multiply IR divergent triangles $I_3^{IR}$ in
Eq.~(\ref{eq:genamp}).   All of the other coefficients that multiply the finite
integrals can be determined numerically using the recursion relations. 
However, because the  coefficient $K^{IR}$ multiplies a divergent integral, we
need to know it in $D$ dimensions. This can only be achieved using analytic
methods. In Sec.~\ref{sec:IR1}, we decompose the divergent part of a rank-$m$
$N$-point function into a sum of tensor structures multiplied by the divergent
triangles and a kinematic factor, $C_N^{12,j}$.  This kinematic factor is the building block 
for the coefficient $K^{IR}$.

First we derive an analytic expression for $C_N^{12,j}$ in
Sec.~\ref{subsec:c123}. As it turns out this coefficient does not depend on the
rank of its parent integral and all the tensor information is carried by a 
kinematic Lorentz structures. 

While in this section we consider only massless internal lines, the 
generalization to include masses is straightforward and has been outlined in
Ref.~\cite{dittmaier}. By combining results from \cite{dittmaier}  with the
techniques to deal with tensor integrals explained in this section, one
readily obtains the extension to the massive case. 

\subsection{Derivation of the Divergent Coefficients}
\label{sec:IR1}

As we saw in Sec.~\ref{sec:notation}, the IR divergences of any (tensor) loop
integral can  be isolated as a combination of triangle integrals with one
and two light-like legs~\cite{binoth,kinoshita}. In the standard approach   
using the recursion relations,  the IR divergent graphs are produced from
many different sources and ultimately drop out at the end of a lengthy
algebraic calculation, often after large cancellations between terms. Here it
is our aim to calculate the kinematic coefficients multiplying these infrared
divergent integrals {\it ab initio} 
without recourse to the recursion relations.  In other
words, starting with some given (tensor) loop integral,  we would like to take
the IR limit directly and isolate the singularities,
\begin{eqnarray}
I_N^{\mu_1\cdots\mu_m}(D;\nu_1,\ldots\nu_N)&
\stackrel{{\rm IR ~limit}} \longrightarrow &S_N^{\mu_1\cdots\mu_m},
\end{eqnarray}
where $S_N$ contains all the infrared singularities.
Ultimately, this will produce exactly the 
correct coefficients $K^{IR}_{\nu_1\nu_2\nu_3}$
that would be produced by application of the recursion relations so that
\begin{eqnarray}
{\cal A}_M^{IR}(p_1,p_2,\ldots,p_M)
&=&\sum_G\sum_{m=0}^{N} {\cal C}^{G}_{\mu_1\mu_2\cdots\mu_m}
S_N^{\mu_1\cdots\mu_m}\nonumber \\
&\equiv&
\sum_{\nu_1\nu_2\nu_3}
K^{IR}_{\nu_1\nu_2\nu_3}I_3^{IR}(D=2(\sigma-1);\nu_1,\nu_2,\nu_3).
\end{eqnarray}
As we will show, identifying $S_N^{\mu_1\cdots\mu_m}$ is straightforward for scalar integrals 
but is 
slightly more involved for tensor integrals.

\subsubsection{Scalar Integrals}

The infrared singularity corresponds to the limit where the incoming loop
momentum at the vertex where particle $i$ joins the graph becomes collinear
with  the incoming momentum $p_i$.  If we define
$\ell_i=\ell+p_1+\cdots+p_{i-1}$ ($\ell_1=\ell$) to be the momentum entering
the vertex and $\ell_i + p_i$ to be the momentum exiting the vertex, then the
collinear limit occurs when $\ell_i \rightarrow xp_i$. Both propagators
adjacent to the incoming momentum $p_i$ are now on-shell, $\ell_i^2 = x^2 p_i^2
= 0$ and $(\ell_i+p_i)^2 \to (1+x)^2p_i^2 = 0$, and produce infrared
singularities. 
We therefore can construct a term that matches all of 
the infrared singularities when the loop momentum becomes collinear
with one of the external momenta. 
With the identification $d_N=d_0$ the counterterm for a scalar $N$-point function
in the limit $\ell_2 \rightarrow xp_2$ is given by
\begin{equation}
\label{eq:S2scalar}
\JJ_{N,2}=\Delta(p_2^2)\int\frac{d^D \ell}{i\pi^{D/2}}\frac{1}{d_1d_2} 
~\left[\frac{1}{d_3(x)d_4(x)\cdots d_N(x)}\right]_{\ell_2\rightarrow xp_2}.
\end{equation} 
where $\Delta(p_2^2)=1$ when $p_2^2=0$ and zero otherwise 
(i.e. when $p_2$ is off-shell).
The square bracket is to be evaluated in the limit that
$\ell_2\rightarrow xp_2$, or equivalently $\ell \rightarrow xq_2-(1+x)q_1$. 
So far, the value of $x$ has not been determined.   
However, there are still singularities 
when $x$ takes on the specific value $x =x_j$ such that $d_j(x_j) = 0$.  
These singularities can be extracted one at a time so that,
\begin{equation}\label{eq:cj}
\JJ_{N,2} = \Delta(p_2^2) \sum_{j=3}^{N}
\DD_{12,j}(1,1,1)~\CC_N^{12,j},
\end{equation} 
where,
\begin{equation}
\DD_{12,j}(1,1,1)=I_3(D=4-2\e;\{q_1,q_2,q_j\},1,1,1)\equiv
\int\frac{d^D \ell}{i\pi^{D/2}}\frac{1}{d_1d_2d_j} 
\end{equation}
is independent of the number of legs in the graph. 
The kinematic factor is given by,
\begin{equation}\label{eq:kinfac}
\CC_N^{12,j}=\frac{1}{\prod_{\stackrel{i=3}{i\neq j}}^Nd_i(x_j)}\ .
\end{equation}
We now have accomplished our aim for the scalar integral. 
The IR singularities associated with
incoming light-like momentum $p_2$ is a sum of kinematic factors
multiplying scalar three-point master integrals. 

\subsubsection{Tensor Integrals}

In the case of tensor integrals the derivation is a bit more complicated.
In the collinear limit $\ell_2\rightarrow xp_2$ we find
\begin{equation}
\label{eq:S2tensor}
\JJ_{N,2}^{\mu_1\cdots\mu_m}=\Delta(p_2^2)\int\frac{d^D \ell}{i\pi^{D/2}}
\left.\frac{\ell^{\mu_1}\cdots\ell^{\mu_{m_3}}}{d_1d_2}
~\left[\frac{\ell^{\mu_{m_3+1}}\cdots\ell^{\mu_k}}
{d_3(x)d_4(x)\cdots d_N(x)}\right]_{\ell_2\rightarrow xp_2}
\right|_{\mbox{IR limit}} 
\end{equation} 
where $m_3$ of the $m$ loop momenta are integrated over and the remaining
$(m-m_3)$ are evaluated in the IR limit $\ell_2\rightarrow xp_2$. 
A priori, any value of $m_3$ between 0 and $m$ is possible and
leads to the correct IR behaviour. In Ref.~\cite{dittmaier} it is
proposed that $m_3$ is chosen to be equal to $m$. Using the recursion
relations of Sec.~\ref{sec:notation} this particular choice would give
a rather complicated structure of IR triangles. However, as explained 
above more generic choices for the tensor structure or the IR terms
are possible. 

To decide on the appropriate partition of loop momenta, 
it is important to consider the structure of the IR
triangles produced by the recursion relations of Sec.~\ref{sec:recursion}.
Once we can get the correct correspondence between
the two different methods for evaluating the IR contributions 
we have by principle of
consistency found the right procedure for partitioning the loop momenta.

\FIGURE[t!]{
\label{fig:sec3}
\scalebox{.5}{\includegraphics{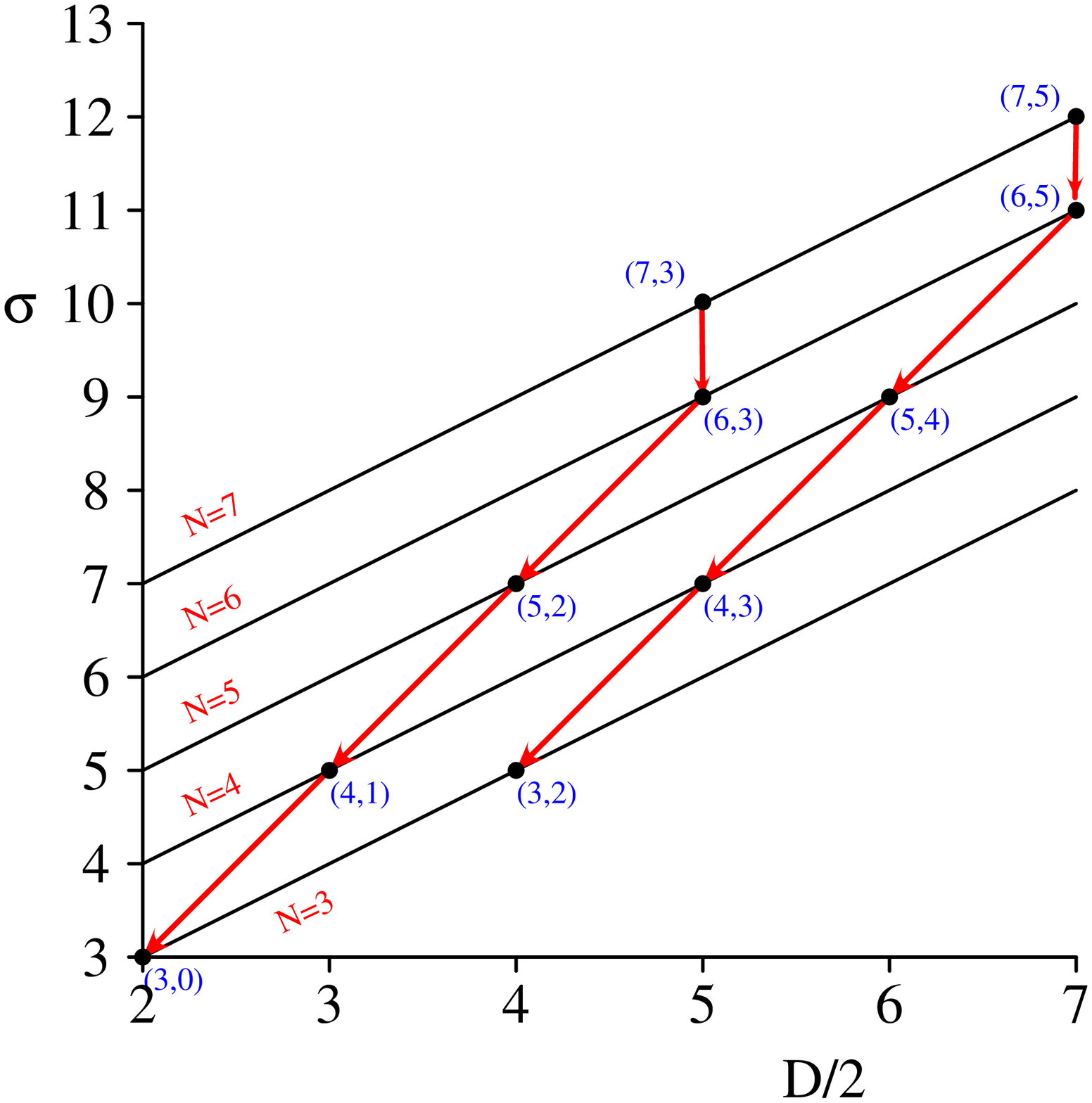}}
\caption{A plot in $(D/2,\sigma)$ showing how the IR divergences from
high-rank $N$-point integrals are mapped onto IR divergent triangle integrals
using Eqs.~(\ref{eq:recursion1}) and (\ref{eq:recursion6}).
The solid lines show the IR trajectories for $N=3,\ldots,7$.
The points labelled $(N,m)$ correspond to integrals produced by rank-$m$
 $N$-point tensor integrals.   The red arrows 
 show how Eqs.~(\ref{eq:recursion1}) and (\ref{eq:recursion6}) 
 systematically reduce 
 both $N$ and $\sigma$ until the integral ends on the $N=3$ IR line.}
}

Fig.~\ref{fig:sec3} shows how the IR divergences from rank-$m$ $N$-point 
integrals are mapped onto IR divergent triangle integrals using
Eqs.~(\ref{eq:recursion1}) and (\ref{eq:recursion6}). 
Because the recursion relations change in a fundamental matter
once $N\geq 7$ we have to distinguish these two cases.

For $N\leq 6$, systematic
application of identity ~(\ref{eq:recursion1})
reduce both $\sigma$  and $D$ in such
a way that the  integral jumps from the $N$-point IR trajectory to the
$(N-1)$-point IR trajectory. Each jump reduces $\sigma$ by 2 and $D/2$
by unity. Repeated application of the identity will always end on
the $N=3$ IR line.   Provided that the rank-$m$ satisfies $m\leq N-
3$, then we obtain  triangle integrals with $\sigma = 3$.   However,
if $m > N-3$, then the recursion relations hit the IR triangle line at
$\sigma > 3$ and $D/2 > 2$.   In Fig.~\ref{fig:sec3} this is
illustrated  for rank-5 six-point integrals  $(6,5)$.   After
reduction, we are left with triangle integrals in $D/2=4$ and with
$\sigma =5$, precisely the integrals that are produced by rank-2
triangle graphs in $D/2=2$.

For $N\geq 7$ the recursion relation  ~(\ref{eq:recursion6}) works 
rather differently. It keeps $D/2$ fixed, but reduces $\sigma$ and $N$.
Repeated application will eventually move the integral onto
the $N=6$ IR line and the previous case is applied.

This leads us to the surprisingly simple conclusion that the correct
partitioning of loop momenta in  Eq.~(\ref{eq:S2tensor}).
For $N\leq 6$ we find $m_3=\max(m+3-N,0)$, while for $N\geq 7$ we have
$m_3=\max(m-3,0)$. Or summarized in one formula
\begin{equation}\label{eq:m3def}
m_3=\max(m+3-\min(N,6),0).
\end{equation}
When $N \leq 6$ the maximum value of
$m_3$ is only three (reached for a maximum rank-$m=N$ $N$-point function).
However, if $N > 6$, the maximum value of $m_3$ is $N-3$.   Note that 
only the IR divergent part of the tensor triangle integral is
required.  

Inserting this value of $m_3$ in Eq.~(\ref{eq:S2tensor}) and 
partitioning the loop momenta in a symmetric way the counterterm 
can be written as
\begin{eqnarray}
\label{eq:S3}
\JJ_{N,2}^{\mu_1\cdots\mu_m}&=&
\Delta(p_2^2)\sum_{j=3}^N \CC_N^{12,j}\TT^{\mu_1\cdots\mu_m}_{12,j},
\end{eqnarray} 
where 
\begin{equation}
\TT^{\mu_1\cdots\mu_m}_{12,j}
=\DD_{12,j}^{[\mu_1\cdots\mu_{m_3}}
\PP_{12,j}^{\mu_{m_3+1}\cdots\mu_m]}.
\end{equation}
The notation
\begin{equation}
A^{\large[\mu_1\cdots\mu_k}B^{\mu_{k+1}\cdots\mu_m\large]}
=\sum_{P(\mu_1\cdots\mu_m)}\frac{1}{k!}\frac{1}{(m-k)!} 
A^{\mu_1\cdots\mu_k}B^{\mu_{k+1}\cdots\mu_m}
\end{equation}
shows how the Lorentz indices are distributed between the terms and
includes a sum over all permutations of the indices.
The kinematic factor $\CC_N^{12,j}$ is defined in Eq.~(\ref{eq:kinfac}) while
the tensor structure remaining in the IR limit is given by,
\begin{eqnarray}
\PP_{12,j}^{\nu_1\cdots\nu_k}&=&\left[\ell^{\nu_1}\cdots\ell^{\nu_k}
\right]_{\ell\rightarrow x_jp_2-p_1}.
\end{eqnarray}
Finally the IR part of the rank-$k$ triangle function is given by,
\begin{eqnarray}\lefteqn{
\DD_{12,j}^{\nu_1\cdots\nu_{k}}=
\sum_{x_1x_2}\left(q_1^{x_1}q_2^{x_2}q_j^{k-x_1-x_2}\right)^{\nu_1\cdots\nu_{k}}}
\nonumber\\
&&\times I_3^{IR}(D=4+k-2\e;q_1,q_2,q_j,x_1+1,x_2+1,k+1-x_1-x_2).
\end{eqnarray}
Eq.~(\ref{eq:S3}) produces the IR singularities 
for the rank-$m$ $N$-point integral in
the  $l_2\rightarrow xp_2$ limit. 
Note that only those 3-point functions  
which are divergent (i.e. have at least one massless on-shell leg) are kept.

We see that Eq.~(\ref{eq:S3})
contains exactly the same IR triangles as would be obtained by
using the formalism of Sec.~\ref{sec:notation} to a
rank-$m$ $N$-point function followed by application of the recursion relations.
Therefore, by inspection, the divergences of the  Eq.~(\ref{eq:S3})
are identical to the original tensor integral.
Because the kinematic factor $\CC_N^{12,j}$ does not contain
any dimensional term (i.e. does not depend on the regulator $\e$),
the finite parts will also match exactly and,
\begin{equation}
\lim_{l_2\rightarrow xp_2}
\left[I_N^{\mu_1\cdots\mu_m}-S_{N,2}^{\mu_1\cdots\mu_m}\right]
\equiv 0.
\end{equation}
 
\subsubsection{Sum over all IR limits}

We have constructed the IR divergent part of the loop integral associated with
the particular limit $\ell_2\rightarrow xp_2$.
Similar IR singular terms for the other cases (i.e. $\ell_i \rightarrow xp_i$) can be
obtained by cyclic  permutations of Eq.~(\ref{eq:cj}). That is
$(\ell_1,p_1,d_1)\rightarrow (\ell_2,p_2,d_2)
\rightarrow\cdots\rightarrow(\ell_N,p_N,d_N)\rightarrow(\ell_1,p_1,d_1)$ denoted by
$C(12\cdots n)$.  However, care has to be taken when adding up all contributions to
construct the overall IR singular term.  As is well known for real emission
soft/collinear divergences, adding up all collinear limits will double count the soft
double singularities. The same occurs for the virtual soft poles when two adjacent
external momenta are on-shell. For example, when both $p_2$ and $p_3$ are on-shell we get an
identical soft/double pole contribution from $\JJ_n^{(2)}$ and $\JJ_n^{(3)}$.
These terms are respectively $\CC_n^{12,3}/d_1d_2d_3$ and $\CC_n^{23,1}/d_2d_3d_1$.
To correct for the double counting we have to subtract this singularity.   With this
in mind we can now construct the overall IR singular term for the 
loop integral,
\begin{eqnarray}
\label{eq:subtraction}
\JJ_N^{\mu_1\cdots\mu_m} &=&
\sum_{C(12\cdots n)} \left[\JJ_{N,2}^{\mu_1\cdots\mu_m}
-\Delta(p_2^2)\Delta(p_3^2)\TT_{12,3}^{\mu_1\cdots\mu_m}\CC_N^{12,3} \right]
\nonumber \\ &=& \sum_{C(12\cdots n)}\left[
\sum_{j=3}^n \Delta(p_2^2)\TT_{12,j}^{\mu_1\cdots\mu_m}\CC_N^{12,j}
- \Delta(p_2^2)\Delta(p_3^2)\TT_{12,3}^{\mu_1\cdots\mu_m}\CC_N^{12,3}
\right]
\nonumber \\ &=& \sum_{C(12\cdots N)}\left[
\sum_{j=4}^{n-1} \Delta(p_2^2)\TT_{12,j}^{\mu_1\cdots\mu_m}\CC_N^{12,j}
\right.\nonumber\\ &&\phantom{ = \sum_{C(12\cdots  N)}}\left.
+\Delta(p_2^2)\Delta(p_3^2)\TT_{12,3}^{\mu_1\cdots\mu_m}\CC_N^{12,3}
\right.\nonumber\\ &&\phantom{ = \sum_{C(12\cdots n)}}\left.
+\Delta(p_2^2)\left(1-\Delta(p_3^2)\right)\TT_{12,3}^{\mu_1\cdots\mu_m}\CC_N^{12,3}
\right.\nonumber \\
&&\phantom{ = \sum_{C(12\cdots N)}}\left.
+\Delta(p_3^2)\left(1-\Delta(p_2^2)\right)\TT_{23,1}^{\mu_1\cdots\mu_m}\CC_N^{23,1}
\right].
\end{eqnarray}
Eq.~(\ref{eq:subtraction}) clearly shows the
different contributions. Just as for real emission soft/collinear
behavior, there is a double pole soft contribution when two neighboring
incoming momenta are on-shell, $\Delta(p_2^2) = \Delta(p_3^2) = 1$. 
If one of the two neighboring incoming
lines is off-shell there is a single pole collinear contribution. However,
unlike the real emission case, there is an additional single pole
contribution from non-adjacent propagators. 

\subsubsection{Evaluation of $\CC_N^{12,j}$}
\label{subsec:c123}
If the momentum $p_2$ is light-like, then the condition that $\ell_2\rightarrow
xp_2$  means that all of the propagator factors become first order
polynomials in $x$.  In this case, 
we can use,
\begin{eqnarray}
  \Pp{\ell_2+p_2+\cdots+p_j}\Big\rfloor_{\ell_2\rightarrow xp_2}
  =(1+x) s_{2\cdots j}-x s_{3\cdots j},
  =A_j (x-x_j)
\end{eqnarray}
with
\begin{eqnarray}
A_j &=& s_{2\cdots j}-s_{3\cdots j}, \label{eq:PolC1}\\
x_j &=& \frac{s_{2\cdots j}}{s_{3\cdots j}-s_{2\cdots j}}, \label{eq:PolC2}\\
s_{k\cdots j}&=& \Pp{p_k+\cdots +p_j}, \label{eq:PolC3}
\end{eqnarray}
so that 
\begin{equation}
\left[ 
{d_3(x)d_4(x)\cdots d_N(x)}\right]_{\ell_2\rightarrow xp_2}
= \prod_{i=3}^{N} A_i (x-x_i).
\end{equation}
The coefficient $\CC_N^{12,j}$ is merely the 
residue evaluated at the singular
point $x=x_j$,
\begin{equation}
\CC_N^{12,j} =
\frac{1}{\prod_{m=3,m\neq j}^{N} A_m (x_j-x_m)}.
\end{equation}
Inserting the definitions of $A_i$ and $x_i$ given in 
Eqs.~(\ref{eq:PolC1}) and (\ref{eq:PolC2}) we find
\begin{equation}\label{eq:Cgeneral}
\CC_N^{12,j} =
\frac{\left(s_{2\cdots j}-s_{3\cdots j}\right)^{N-3}}{\prod_{k=3,k\neq j}^{N} 
\left(s_{2\cdots j}s_{3\cdots k}-s_{3\cdots j}s_{2\cdots k}\right)}
\end{equation}
where $q_i=\sum_{j=1}^i p_i$ such that the formula respects cyclic permutations
of the indices.
Eq.~(\ref{eq:Cgeneral}) holds whether or not $p_j^2=0$.

The special cases associated with neighboring incoming momenta are 
obtained by setting $j=3$ in Eq.~(\ref{eq:Cgeneral}) and noting that
$s_{3\cdots3} = p_3^2$ so that, 
\begin{equation}
\CC_N^{12,3} =
\frac{\left(s_{23}-p_3^2\right)^{N-3}}{\prod_{k=4}^N 
\left(s_{23}s_{3\cdots k}-p_3^2s_{2\cdots k}\right)},
\end{equation}
corresponding to evaluating $\ell\rightarrow
x_3 p_2-q_1$ with $x_3=s_{23}/(s_{23}-p_3^2)$ and $q_1 = p_1$. 
The companion   
coefficient is given by making the cyclic momentum permutation, $p_1 \to p_2, p_2 \to
p_3, q_1 \to q_2$ on Eq.~(\ref{eq:Cgeneral}) so that $s_{2\cdots j} \to
s_{3\cdots 1} = p_2^2$, $s_{3\cdots j} \to s_{4\cdots 1} = s_{23}$,
\begin{equation}\label{eq:123}
\CC_N^{23,1} =
\frac{\left(s_{23}-p_2^2\right)^{N-3}}{\prod_{k=4}^N 
\left(s_{23}s_{3\cdots k}-p_2^2s_{4\cdots k}\right)},
\end{equation}
corresponding to $\ell \to  x_1 p_3-p_1-p_2$ with 
$x_1=-p_2^2/(p_2^2-s_{23})$.

When $p_2^2=p_3^2=0$ the two coefficients become identical
\begin{equation}
\CC_N^{12,3}\Big\rfloor_{p_2^2=p_3^2=0} 
=\CC_N^{23,1}\Big\rfloor_{p_2^2=p_3^2=0} =
\frac{1}{\prod_{k=4}^N s_{3\cdots k}}.
\end{equation}

As an application of Eq.~(\ref{eq:subtraction}), 
in Appendix~\ref{app:scalar} we write down an expression for the
IR singularities for a scalar massless $N$-point function with
all external momenta on-shell.

\section{Outlook and Conclusions}
\label{sec:conclusions}

The formalism outlined in this paper offers a realistic prospect of
constructing NLO Monte Carlo programs  for processes involving high
multiplicity final state particles. 
The key result is a strategy for
combining the analytic evaluation of the IR
and UV singular contributions to loop integrals by using a suitable basis of
divergent integrals with a system of recursion relations which allows the
numerical determination  of the finite integrals.   This mixed approach 
avoids the
usual algebraic log-jam in loop calculations that occurs when many particles
and kinematic scales are present.   In this paper, we have focussed on 
massless
particles circulating in the loop.   However, the extension to massive
particles is straightforward and opens up the
possibility of computing the NLO corrections to any Standard Model process.

The next phase of this project is the actual numerical implementation of
the algorithms such that a rank-$m$ $N$-point function can be 
calculated. There are two avenues of approach. 
The first approach is to use the
recursive algorithms numerically. The advantages are simplicity and 
all kinematic coefficients are calculated numerically. 
This allows us to check that e.g
$K_5=0$ and all $K^{IR}$ can be compared to the analytic calculation
of these divergent coefficients. The recursive algorithm is a good
diagnostic tool.

However, if the aim is to calculate a
full amplitude made up from many graphs containing different integrals of
varying rank, the recursive approach would not be optimal. This is because
the calculation
of a rank-$m$ $N$-point function uses some of the calculations needed
for evaluating a rank-$k$ $M$-point function ($k\leq m$, $M\leq N$).
%Furthermore, a lot of numerical effort may be spent calculating the 
%kinematic coefficients
%$K_5$, $K^{UV}$ and $K^{IR}$. 
To avoid duplicate calculations and
irrelevant paths, one could use a constructive method for evaluating the
integrals. The algorithm starts by calculating all
the relevant finite basis integrals. Then, step by step, we calculate
the integrals with higher $D$, $N$ and $\sigma$ using the recursion
relations in the reverse sense.
For example, assume all $I_M(D;\{\nu_l\})$ are known 
with $\sigma \leq s_0$ 
(i.e. each integral is a number in a look-up table). 
Using the recursion relations of Sec. 2, we
can then calculate all the $I_N(D;\{\nu_l\})$ with $\sigma=s_0+1$. Now
we know the look-up table for all integrals with $\sigma\leq s_0+1$.
We continue with this procedure until all 
integrals with $M\leq N$ are known. 
After that, building the amplitudes is easy. All the tensor
structures from the Davydychev decomposition are contracted in with
the tree level currents. We can simply look up the numerical value
for the appropriate integral
$I_N(D;\{\nu_l\})$ in the table. This method is very
efficient. However, some of the diagnostic tools have been lost.
(E.g. we assume $K_5=0$ and all divergent coefficients are correct).

Taken together, the combination of recursive numerical and analytic algorithms
should provide a compact and efficient way of evaluating multiparticle 
one-loop amplitudes, thereby opening up the possibility of
estimating
the NLO QCD corrections  for processes such as $pp\rightarrow N$ jets in
association  with vector-boson(s) or Higgs particles. Or for example,
$pp\rightarrow Q\bar{Q}$  plus $N$ jets at NLO, $pp\rightarrow
Q_1\bar{Q_1}\,+\,Q_2\bar{Q_2}\,$ plus $N$ jets at NLO, etc. The only 
limiting
factor on the particles is that the spin/helicity is less than or equal 
to one
and (numerical) computing power. We expect that this latter problem will be
solved with the advent of the LHC Computing Grid. 

\section*{Acknowledgements}

We thank J. Andersen, T. Binoth, T. Birthwright, K. Ellis, G. Heinrich, Z. Nagy,
C. Oleari and G. Passarino for useful discussions.  We thank Giulia Zanderighi
for pointing out a mistake in eq. (2.27) which has been corrected in this
version of the paper.
We also thank the organisers of the
Institute for Particle Physics Phenomenology Monte  Carlo at Hadron Collider
workshop where this work was initiated and the Kavli Institute for Theoretical
Physics  where part of this research was performed. This work was supported in
part by the UK Particle Physics and Astronomy  Research Council and in part by
the EU Fifth Framework Programme  `Improving Human Potential', Research
Training Network `Particle Physics Phenomenology  at High Energy Colliders',
contract HPRN-CT-2000-00149.

\newpage
\appendix

\section{Derivation of Recursion Relations}
\label{app:recursion}

The one-loop recursion relations are based on the integration by parts techniques
developed in Ref.~\cite{Chetyrkin} for calculating higher order beta functions.
The application to one-loop multi-leg recursion relations was developed in
Ref.~\cite{looprecur,Nizic}. Here we will initially follow Ref.~\cite{Nizic} by
applying the integration by parts identity to
\begin{equation} 
\int\frac{d^D\ell}{(2\pi)^D}\frac{\partial}{\partial\ell^{\mu}}
\left(\frac{\left(\sum_{i=1}^N y_i\right)\ell^\mu+\left(\sum_{i=1}^N y_iq_i^\mu\right)}
{d_1^{\nu_1}d_2^{\nu_2}\cdots d_N^{\nu_N}}\right).
\end{equation} 
After some trivial algebra and applying the dimensional shift identity
\begin{equation}\label{eq:recurId}
I_N(D-2;\{\nu_k\})=-\sum_{i=1}^N\nu_i\,I_N(D;\{\nu_k+\delta_{ik}\})\ ,
\end{equation} 
one finds the base equation
\begin{eqnarray}
\label{eq:base}
\lefteqn{
  \sum_{j=1}^N\left(\sum_{i=1}^N S_{ji} y_i\right)\nu_j 
  I_N(D;\{\nu_l+\delta_{lj}\})} \nonumber \\
&=&-\sum_{i=1}^N y_i I_N(D-2;\{\nu_l-\delta_{li}\})
-\left(D-1-\sum_{j=1}^N\nu_j\right)\left(\sum_{i=1}^N y_i\right)
I_N(D;\{\nu_l\})\ ,
\end{eqnarray}
which is valid for any choice of the $N$ parameters $\{y_i\}$. 

To reduce tensor integrals in a systematic way,  more general recursion relations
are needed than those previously given in the literature~\cite{looprecur,Nizic}.
For any useful identity, the parameters
$\{y_i\}$ must be chosen such that the matrix expression 
$\sum_{i=1}^N S_{ji} y_i$
is either a delta function, zero or a unit vector. In all these cases,
the left hand side of Eq.~(\ref{eq:base}) reduces to a single term thereby
yielding a useful recursion relation. 

Note that while the recursion relation is expressed in terms of the
kinematic matrix $S$, the underlying Gram matrix $G$, also 
plays an important role.
For example, for 
$N\geq 6$ the Gram determinant vanishes and this has generic consequences
for the parameters $\{y_i\}$. To understand this we decompose the 
kinematic matrix. The equation to be solved is~\cite{binoth},
\begin{equation}
\sum_{j=1}^N S_{ij}y_j=c\times(1,1,\ldots,1)\equiv\sum_{j=1}^N
\left(\frac{1}{2}G_{ii}y_j+\frac{1}{2}G_{jj}y_j-G_{ij}y_j\right)
\end{equation}
for both $c=0$ and $c=1$. 
In other words, 
\begin{equation}
\sum_{j=1}^{N-1} G_{ij}y_j=\frac{1}{2}Bw_i,\qquad\qquad
\sum_{j=1}^{N-1} w_j y_j=c ,\qquad\qquad
\sum_{j=1}^N y_j=B
\end{equation}
where $w_i=\frac{1}{2}G_{ii}$.

Using momentum conservation at the level of the loop integral, 
we can enforce $B= {\rm det} (G)=0$ by setting $y_N=-\sum_{j=1}^{N-1} y_j$. 
In fact, this particular choice guarantees that 
a solution
exists and that,
\begin{equation}\label{eq:Gramsolution}
\sum_{j=1}^{N-1} G_{ij}y_j\equiv 0,\qquad\qquad
\sum_{j=1}^{N-1} w_jy_j=c, \qquad\qquad
B=\sum_{j=1}^N y_j\equiv 0.
\end{equation}
Therefore any vector in the kernel of $G$ satisfying $w.y=c$ {\em is}  
a solution of $(S.y)_i=c$ for $N\geq 6$.

We can now construct the specific $\{y_i\}$ for all the cases specified in
sec.~\ref{sec:recursion}. To do so we apply the Singular Value Decomposition
(SVD) technique. Using the decomposition we can explicitly
construct the parameters $\{y_i\}$ for Eqs.~(\ref{eq:recursion1}),
(\ref{eq:recursion5}) and (\ref{eq:recursion6}). Note that Eq.~(\ref{eq:recursion2}) follows
from Eq.~(\ref{eq:recursion1}) by summing over index $k$ and applying identity (\ref{eq:recurId}).
Subsequently, Eq.~(\ref{eq:recursion3}) follows by combining Eqs.~(\ref{eq:recursion2}) and 
(\ref{eq:recursion1}).

The three distinct situations are:\\

\noindent{{\bf I. $N\leq 6$}}: $\det(S)\neq 0$, $\sum_{j=1}^N
S_{ij}b_j=\delta_{ik}$ \newline

In this case the inverse of the kinematic matrix exists and the range of
the matrix covers all of parameter space, specifically the unit vectors. 

While many algorithms exist for calculating the inverse, we still use
the stable SVD for the kinematic matrix. This gives good control over the numerical
accuracy of the matrix inversion which will be important 
as the external momenta approach exceptional configurations. The SVD is given by
\begin{equation}
S_{ij}=\sum_{k=1}^N \omega_k u_{ik}v_{kj}^T,
\end{equation} 
where the $N$ $\omega_k$ are the non-zero eigenvalues of $S$ and
the matrices $u$ and $v$ are orthogonal i.e. $(v.v^T)_{ij}=(u.u^T)_{ij}=\delta_{ij}$.
For an explicit algorithm which calculates both the eigenvalues and the orthogonal
matrices see e.g.~\cite{numeric}.

For any $k\leq N$, the choice,
\begin{equation}
b_i=S_{ik}^{-1}=\sum_{j=1}^N \omega_j^{-1}v_{ij}u_{jk}^T,
\end{equation}
yields $S_{ij}b_j=\delta_{ik}$ which immediately leads to Eq.~(\ref{eq:recursion1}). 
Note that because of Eq.~(\ref{eq:Gramsolution}),
when $N=6$ we find the special identity
$B=\sum_{ij} S_{ij}^{-1}=0$, 
which then leads directly to Eq.~(\ref{eq:recursion4}). \\

\noindent{{\bf II. $N\geq 7$}}: $\det(S)=0$, $\sum_{j=1}^N S_{ij}z_j=0$
\newline

In this case the inverse is no longer defined. However for any vector in
the kernel of the singular matrix $S$ we have $\sum_j S_{ij}z_j=0$. By 
selecting $z_i$ with the property $\sum_i z_i=0$, we find Eq.~(\ref{eq:recursion5}). 

To construct this solution,we simply solve Eq.~(\ref{eq:Gramsolution})
by picking a vector out of the kernel of $G$ such that $w.z=0$
For explicit construction
of these vectors we first apply the SVD to the Gram matrix (and {\it not}
the kinematic matrix),
\begin{equation}\label{eq:gram}
G_{ij}=\sum_{k=1}^4 \omega_k u_{ik}v_{kj}^T.
\end{equation} 
With this decomposition, one of the possible choices for $z_i$ is
\begin{equation}
z_i=\frac{W_6 v_{i5}-W_5 v_{i6}}{W_5+W_6},
\end{equation}
where
\begin{equation}
W_i=\frac{1}{2}\sum_{j=1}^{N-1} G_{jj} v_{ji}.
\end{equation}
Note that this solutions has the special property $B=\sum_{i=1}^N z_i=0$ by 
construction.\\

\noindent{{\bf III. $N\geq 7$}}: $\det(S)=0$, $\sum_j S_{ij}r_j=1$  \newline

The fact that the vector $(1,1,\ldots,1)$ is in the range of the singular matrix 
is a highly non-trivial statement. It is due to the properties of the underlying
decay/scattering process. With the property that $\sum_j S_{ij}r_j=1$ and
$\sum_j r_j=0$ we readily obtain Eq.~(\ref{eq:recursion6}).

By inspecting Eq.~(\ref{eq:Gramsolution}), we know that a solution exists.
We simply pick a vector out of the kernel of $G$ such that $w.r=0$.

A explicit construction of the parameters $r_i$ again passes through the SVD of the
Gram matrix, Eq.~(\ref{eq:gram}). An example of the $r_i$ satisfying all of the
above requirements is given by,
\begin{equation}
r_i=\frac{v_{i5}}{W_5}\ .
\end{equation}
\\

With these explicit constructions of the solutions for $(b,z,r)$, we have fully 
specified the recursion relations. The SVD yields, not only an explicit
construction of the solution, but also a diagnostics tool for exceptional momentum
configurations. These configurations can occur, even in non-singular cases. While
integrable, they may require special treatment to maintain numerical accuracy.
The SVD is particular useful to identify these situations.
  
\section{Analytic forms for the divergent integrals}
\label{app:analytic}

The one-loop bubble integral with external momentum scale $Q_1^2$ 
with arbitrary powers of massless propagators is given by
\begin{equation}
I_2(D; \{\nu_1,\nu_2\}) = (-1)^{\sigma}(-Q_1^2)^{\halfD-\sigma}
\frac{\G{\halfD-\nu_1}\G{\halfD-\nu_2}\G{\sigma-\halfD}}
{\G{ \nu_1}\G{ \nu_2}\G{D-\sigma}}
\end{equation}
Inserting specific choices for $\{\nu_1,\nu_2\}$, we see that,
\begin{eqnarray}
I_2(D=4-2\e;\{1,1\}) &=& c_\Gamma \frac{1}{\e(1-2\e)}\left(-Q_1^2\right)^{-\e},\\
I_2(D=6-2\e;\{1,1\}) &=& c_\Gamma \frac{1}{2\e(1-2\e)(3-2\e)}\left(-Q_1^2\right)^{1-\e},\\
I_2(D=6-2\e;\{2,1\}) &=& -c_\Gamma \frac{1}{2\e(1-2\e)}\left(-Q_1^2\right)^{-\e},\\
I_2(D=8-2\e;\{2,2\}) &=& c_\Gamma \frac{(1-\e)}{2\e(1-2\e)(3-2\e)}\left(-Q_1^2\right)^{-\e},\\
I_2(D=8-2\e;\{3,1\}) &=& c_\Gamma \frac{(2-\e)}{4\e(1-2\e)(3-2\e)}\left(-Q_1^2\right)^{-\e}.
\end{eqnarray}
where
\begin{equation}
c_\Gamma = \frac{\Gamma(1-\e)^2\Gamma(1+\e)}{\Gamma(1-2\e)}.
\end{equation}

The one-loop triangle integral with two on-shell legs, 
one off-shell external momentum scale $Q_1^2$ 
and with arbitrary powers of massless propagators is given by
\begin{eqnarray}
I_3(D;\{\nu_1,\nu_2,\nu_3\})\!\!
&=& \!\!(-1)^{\sigma} (-Q_1^2)^{\halfD-\sigma} \nonumber\\
&\times & \!\!
\frac{\G{\halfD-\nu_1-\nu_2}\G{\halfD-\nu_1-\nu_3}\G{\sigma-\halfD}}
{\G{\nu_2}\G{\nu_3}\G{D-\sigma}}.
\label{eq:tri1result}
\end{eqnarray}
Note that $\nu_1$ refers to the propagator opposite the off-shell 
external leg.

The one-loop triangle integral with one on-shell legs, 
two off-shell external momentum scales $Q_1^2 > Q_2^2$ 
and with arbitrary powers of massless propagators is given by

\begin{eqnarray}
\lefteqn{I_3(D; \{\nu_1,\nu_2,\nu_3\})}\\ 
&=&(-1)^{\sigma}
   ~(-Q_1^2)^{\frac{D}{2}-\sigma}
\frac{\Gamma\left(\frac{D}{2}-\nu_1-\nu_2\right)
\Gamma(\frac{D}{2}-\nu_1-\nu_3)\Gamma(\sigma-\frac{D}{2})}
{\Gamma(\nu_2)\Gamma(\nu_3)\Gamma(D-\sigma)}
\nonumber \\
&& 
\hspace{3cm}
\times 
\phantom{~}{_2F_1}\left(\nu_1,\sigma-\frac{D}{2},
1+\nu_1+\nu_3-\frac{D}{2},\frac{Q_2^2}{Q_1^2}\right)
      \nonumber \\
&+&(-1)^{\sigma}
   ~(-Q_1^2)^{-\nu_2}
   ~(-Q_2^2)^{\frac{D}{2}-\nu_1-\nu_3}
\frac{\Gamma\left(\frac{D}{2}-\nu_1-\nu_2\right)
\Gamma\left(\frac{D}{2}-\nu_3\right)
\Gamma\left(\nu_1+\nu_3-\frac{D}{2}\right)}
{\Gamma(\nu_1)\Gamma(\nu_3)\Gamma(D-\sigma)}
\nonumber \\
&& 
\hspace{3cm} \times
\phantom{~}{_2F_1}\left(\nu_2,\frac{D}{2}-\nu_3,
               1+\frac{D}{2}-\nu_1-\nu_3,\frac{Q_2^2}{Q_1^2}\right)
~. \phantom{aaa}
\label{eq:tri2result} 
\end{eqnarray}
Note that $\nu_3$ refers to the propagator opposite the on-shell 
external leg.

\section{The scalar $N$-point function}

\label{app:scalar}
Using Eq.~(\ref{eq:subtraction}) 
can write down the singular structure for
the scalar $N$-point function with all legs on-shell,
\begin{equation}\label{eq:cjn}
\JJ_N =\sum_{C(12\cdots N)}\left[
-\DD_{12,3} \CC_N^{12,3} 
+\sum_{j=3}^N\DD_{12,j} \CC_N^{12,j} 
\right].
\end{equation}
Integrating out the loop momentum in the scalar triangle integrals
yields 
\begin{eqnarray}
\JJ_N&=&c_\Gamma\frac{1}{\e^2}\sum_{C(12\cdots n)}
\left[
-\frac{(-s_{23})^{-\e}}{\prod_{k=4}^N s_{2\cdots k}}
\right.\nonumber\\&&\left.\phantom{ }
+\sum_{j=3}^N\frac{(s_{23\cdots j}-s_{3\cdots j})^{N-4}}
{\prod_{k=3,k\neq j}^N(s_{23\cdots j}s_{3\cdots k}-s_{3\cdots j}s_{23\cdots k})}
\left((-s_{23\cdots j})^{-\e}-(-s_{3\cdots j})^{-\e}\right)
\right].
\end{eqnarray}
This expression agrees with the singular structure singularity
structure for the massless six-point function with all momenta on-shell
given in Ref.~\cite{binoth} and the corresponding 
massless five-point function with all momenta on-shell given 
in Ref.~\cite{Apentagon}
as well as the massless four-point function of Ref.~\cite{Abox}.

It is straightforward to use Eq.~(\ref{eq:subtraction}) to check the known
singularity structure for  massless box graphs with all possible combinations of
off-shell conditions  for the external momenta given in Ref.~\cite{Abox} as well as
the massless pentagon with one off-shell external momenta~\cite{Apentagon}.
  
\newpage
\bibliographystyle{JHEP}

\end{document}